%% file: ANA-BPHY-2018-10-PAPER.tex
\pdfoutput=1
\documentclass[PAPER, REPORT=false, biblatex=true, atlasdraft=false, texlive=2023, UKenglish, cernpreprint , texmf, orcidlogo]{atlasdoc}
\let\chapter\section\newenvironment{DIFnomarkup}{}{}
\usepackage{datatool}
\usepackage{filecontents}
\usepackage{amsmath}
\usepackage{lscape}
\usepackage{enumitem}
\usepackage{atlaspackage}
\usepackage{placeins}
\usepackage{rotating}

\usepackage{dcolumn}

\usepackage{array}
\usepackage{arydshln}
\usepackage{atlasbiblatex}

\usepackage{atlascontribute}
\usepackage{csvsimple}
\usepackage{atlasphysics}

\usepackage{atlasjetetmiss}
\graphicspath{{logos/}{figures/}{figures/*/}}

\usepackage{ANA-BPHY-2018-10-INT1-defs}

\AtlasTitle{Differential cross-section measurements of $D^{\pm}$ and $D_{s}^{\pm}$ meson production in proton--proton collisions at $\sqrt{s} = \SI{13}{TeV}$ with the ATLAS detector}

\AtlasVersion{3.4       }

\AtlasAbstract{
The production of \(D^{\pm}\) and \(D_{s}^{\pm}\) charmed mesons is measured using the \(D^{\pm}/D_{s}^{\pm} \to \phi(\mu\mu)\pi^{\pm}\) decay channel with 137 fb\(^{-1}\) of \(\sqrt{s} = 13\) TeV proton--proton collision data collected with the ATLAS detector at the Large Hadron Collider during the years 2016--2018. The charmed mesons are reconstructed in the range of transverse momentum \(12 < p_{T} < 100\) GeV and pseudorapidity \(|\eta| < 2.5\). The differential cross-sections are measured as a function of transverse momentum and pseudorapidity, respectively, and compared with next-to-leading-order QCD predictions. The predictions are found to be consistent with the measurements in the visible kinematic region within the large theoretical uncertainties.
}
\AtlasJournal{JHEP }
\AtlasJournalRef{JHEP 07 (2025) 86}
\AtlasDOI{10.1007/JHEP07(2025)086}
\PreprintIdNumber{CERN-EP-2024-322}

\hypersetup{pdftitle={ATLAS document},pdfauthor={The ATLAS Collaboration}}

\begin{document}

\maketitle
\tableofcontents

\clearpage
%

\section{Introduction}

\newcommand{\DmesonFootnote}{%
In this paper, the $\pm$ symbol is used to represent both charge conjugates inclusively for different $D$ mesons; the $D^0$ symbol is used to represent $D^0$ and $\bar{D}^0$ mesons inclusively.
}

The production of heavy hadrons in proton--proton collisions is a fundamental process that provides a crucial test of perturbative quantum chromodynamics (QCD) calculations.
However, large uncertainties persist in current theoretical predictions due to the fact that the masses of heavy quarks are comparable to the typical energy scales of the hard scattering processes~\cite{PhysRevLett.96.012001,Kniehl:2012ti}.
Moreover, heavy hadrons can be produced promptly via the hadronisation of charm quarks from the initial hard scattering process, or non-promptly in decays of $b$-hadrons; precise predictions of such processes are challenging, due to difficulties in modelling non-perturbative effects such as hadronisation, as well as uncertainties in the fragmentation functions and decay dynamics of heavy-flavour hadrons.
Given the sizeable theoretical uncertainties, experimental constraints on heavy hadron production cross-sections are important to improve calculation techniques, as well as in searches for new physics phenomena, where heavy hadron production is often either a signal process or a significant background process.
For instance, several charmed and bottom mesons have significant decay branching ratios to $\tau$-lepton final states~\cite{ParticleDataGroup:2024cfk}; precise cross-section measurements are necessary for using such decays to search for lepton-flavour-violating $\tau$-lepton decays~\cite{ATL-PHYS-PUB-2018-032} at the Large Hadron Collider (LHC)~\cite{Evans:2008zzb}.

At the LHC, prompt and non-prompt $D$ meson\footnote{\DmesonFootnote} production in proton--proton (\pp) collisions has been measured by different experiments.
The ALICE Collaboration reported the differential production cross-sections of prompt $D^0$, $\D$, $D^{*\pm}$, and $\Ds$ mesons at a centre-of-mass energy of \sqrtsthirteen, in the mid-rapidity region and in different ranges of transverse momentum, \pt; in particular, the \D meson cross-section was measured up to \pt of \SI{50}{\GeV} and the \Ds meson cross-section was measured up to \pt of \SI{36}{\GeV}~\cite{ALICE:2023sgl}.
Recently, the ALICE Collaboration published differential production cross-sections of non-prompt $D^0$, $\D$ and $\Ds$ mesons in the range of $\pt<24$ \SI{}{\GeV}; the results are compared to the prompt measurements to evaluate the non-prompt production fraction~\cite{ALICE:2024xln}. Previously, the ALICE Collaboration also published measurements of $D$ meson production at \sqrttwofiveseven~\cite{ALICE:2012inj, ALICE:2019nxm, ALICE:2017olh, ALICE:2021mgk}.
A study conducted by the CMS Collaboration reported the differential cross-sections of $D^{*\pm}$, $D^0$, and \D with \pt up to \SI{100}{\GeV} based on a partial data sample at \sqrtsthirteen collected in 2016~\cite{CMS-BPH-18-003}.
The LHCb Collaboration published measurements of the production cross-sections of prompt $D^0$, $\D$, $D^{*\pm}$, and $\Ds$ mesons in the range of $\pt<15$ \SI{}{\GeV} in the forward region using \sqrtsthirteen data~\cite{LHCb:2015swx}. Similarly, the LHCb Collaboration published measurements of $D$ meson production at \sqrtfiveseven~\cite{LHCb:2016ikn,LHCb:2013xam}.
The ATLAS Collaboration has measured $D^{*\pm}$, \D, and \Ds meson production at \sqrtseven ~\cite{BPHY-2013-04}, but no such measurement has been performed at \sqrtsthirteen, and the differential cross-section of the \Ds meson has not been measured.

This paper presents a measurement of \D and \Ds meson differential production cross-sections performed with the ATLAS detector using the \pp collision data at \sqrtsthirteen collected during the years 2016--2018.
To measure the \D and \Ds mesons produced at the LHC, this study makes use of decay channels $\D/\Ds\to\phi(\mu\mu)\pi^\pm$ with semileptonic final states.
In comparison to reconstructing the fully hadronic decays of the $D$ mesons, the selected decay channels make use of the precise muon reconstruction and identification in the ATLAS detector, allowing a much cleaner signature with a lower background level.
The final state of two muons and one pion ($\mu\mu\pi$) is fully reconstructed and identified, to construct the $D$ meson decay vertex.
The yields of the \D and \Ds meson signals are then extracted simultaneously by fitting the invariant mass distributions of the $\mu\mu\pi$ system.

To calculate the production cross-sections, simulated events are used to evaluate the reconstruction efficiencies and acceptance.
Because of the difficulties in separating the $D$ mesons produced by an initial charm quark (prompt) and those produced by an initial bottom quark (non-prompt), this study does not attempt to extract the prompt and non-prompt signal yields separately; however, their relative contributions in simulated events are constrained to match the data by fitting the pseudo-proper lifetime distribution.
The yields and the cross-sections presented are inclusive of both prompt and non-prompt $D$ mesons, unless otherwise specified.

The inclusive and differential cross-sections of the \D and \Ds mesons are reported in the range of $|\eta|<2.5$ and $\SI{12}{} < \pT < \SI{100}{\GeV}$ and compared with state-of-the-art next-to-leading-order (NLO) calculations.
This is the first differential measurement of the \Ds meson production reported by the ATLAS Collaboration, and the first time such a measurement has been performed up to transverse momenta of \SI{100}{\GeV} at the LHC.


%
\newcommand{\AtlasCoordFootnote}{%
ATLAS uses a right-handed coordinate system with its origin at the nominal interaction point (IP)
in the centre of the detector and the \(z\)-axis along the beam pipe.
The \(x\)-axis points from the IP to the centre of the LHC ring,
and the \(y\)-axis points upwards.
Polar coordinates \((r,\phi)\) are used in the transverse plane,
\(\phi\) being the azimuthal angle around the \(z\)-axis.
The pseudorapidity is defined in terms of the polar angle \(\theta\) as \(\eta = -\ln \tan(\theta/2)\) and is equal to the rapidity
$ y = \frac{1}{2} \ln \left( \frac{E + p_z}{E - p_z} \right) $ in the relativistic limit.
Angular distance is measured in units of \(\Delta R \equiv \sqrt{(\Delta y)^{2} + (\Delta\phi)^{2}}\).}

\section{ATLAS detector}

The ATLAS detector~\cite{PERF-2007-01} at the LHC covers nearly the entire solid angle around the collision point.\footnote{\AtlasCoordFootnote}
It consists of an inner tracking detector surrounded by a thin superconducting solenoid, electromagnetic and hadronic calorimeters,
and a muon spectrometer incorporating three large superconducting air-core toroidal magnets.

The inner-detector system (ID) is immersed in a \qty{2}{\tesla} axial magnetic field
and provides charged-particle tracking in the range \(|\eta| < 2.5\).
The high-granularity silicon pixel detector covers the vertex region and typically provides four measurements per track,
the first hit generally being in the insertable B-layer (IBL) installed before Run~2~\cite{ATLAS-TDR-19,PIX-2018-001}.
It is followed by the SemiConductor Tracker (SCT), which usually provides eight measurements per track.
These silicon detectors are complemented by the transition radiation tracker (TRT),
which enables radially extended track reconstruction up to \(|\eta| = 2.0\).
The TRT also provides electron identification information
based on the fraction of hits (typically 30 in total) above a higher energy-deposit threshold corresponding to transition radiation.

The calorimeter system covers the pseudorapidity range \(|\eta| < 4.9\).
Within the region \(|\eta|< 3.2\), electromagnetic calorimetry is provided by barrel and
endcap high-granularity lead/liquid-argon (LAr) calorimeters,
with an additional thin LAr presampler covering \(|\eta| < 1.8\)
to correct for energy loss in material upstream of the calorimeters.
Hadronic calorimetry is provided by the steel/scintillator-tile calorimeter,
segmented into three barrel structures within \(|\eta| < 1.7\), and two copper/LAr hadronic endcap calorimeters.
The solid angle coverage is completed with forward copper/LAr and tungsten/LAr calorimeter modules
optimised for electromagnetic and hadronic energy measurements respectively.

The muon spectrometer (MS) comprises separate trigger and
high-precision tracking chambers measuring the deflection of muons in a magnetic field generated by the superconducting air-core toroidal magnets.
The field integral of the toroids ranges between \num{2.0} and \qty{6.0}{\tesla\metre}
across most of the detector.
Three layers of precision chambers, each consisting of layers of monitored drift tubes, cover the region \(|\eta| < 2.7\),
complemented by cathode-strip chambers in the forward region, where the background is highest.
The muon trigger system covers the range \(|\eta| < 2.4\) with resistive-plate chambers in the barrel, and thin-gap chambers in the endcap regions.

The luminosity is measured mainly by the LUCID--2~\cite{LUCID2} detector that records Cherenkov light produced in the quartz windows of photomultipliers located close to the beampipe.

Events are selected by the first-level trigger system implemented in custom hardware,
followed by selections made by algorithms implemented in software in the high-level trigger~\cite{TRIG-2016-01}.
The first-level trigger accepts events from the \qty{40}{\MHz} bunch crossings at a rate below \qty{100}{\kHz},
which the high-level trigger further reduces in order to record complete events to disk at about \qty{1}{\kHz}.

A software suite~\cite{SOFT-2022-02} is used in data simulation, in the reconstruction
and analysis of real and simulated data, in detector operations, and in the trigger and data acquisition
systems of the experiment.


\section{Data sample, event simulation and theoretical predictions}

\newcommand{\Fragmentation}{
In this paper, the fragmentation fractions obtained by averaging the LEP measurements are used, because the \pt range of interest is relatively high compared to the \D and \Ds meson masses. It is also noted that the ALICE Collaboration has measured the fragmentation fractions in the low-\pt range, where the \pt of the quarks is comparable to the mass of the hadrons (${\pt}_Q\approx{m}_{Q}$). If the fragmentation values obtained by the ALICE Collaboration were used, the FONLL prediction would be lowered by approximately 10\%.
}

The data used in this analysis were collected in 2016--2018 with the ATLAS detector in \pp collisions at \sqsthirt at the LHC\@. Data collected during 2015 were not used for this study, because of the unavailability of low-\pt di-muon triggers in the di-muon invariant mass range of interest. The data are selected after requiring each detector component to be fully operational~\cite{DAPR-2018-01}. The analysed data sample corresponds to an integrated luminosity of \lumi ~\cite{DAPR-2021-01}.

To model inelastic events produced in \pp collisions, a large sample of Monte Carlo (MC) simulated events was prepared using the \PYTHIA[8.212]~\cite{Sjostrand:2014zea} event generator. The simulation was performed using leading-order (LO) matrix elements for all $2\rightarrow2$ QCD processes. Initial-state and final-state parton showering were used to simulate the effect of higher-order processes. The \textsc{NNPDF2.3lo}~\cite{Ball:2012cx} parameterisation provided the parton distribution functions (PDF) of the proton. The charm quark and bottom quark masses were set to \SI{1.5}{\GeV} and \SI{4.8}{\GeV}, respectively. The event sample was simulated using the ATLAS A14 set of tuned parameters~\cite{ATL-PHYS-PUB-2014-021}. Separate samples were generated for $\pp\rightarrow c\bar{c}$ and $\pp\rightarrow b\bar{b}$ processes, corresponding to prompt and non-prompt production of $D$ mesons, respectively. The generated \D and \Ds mesons were decayed into $\phi(\mu\mu)\pi^\pm$.
The generated events were passed through a full ATLAS detector simulation~\cite{SOFT-2010-01} based on \GEANT~\cite{Agostinelli:2002hh} and processed with the same reconstruction algorithms as used for the data. The generation of the simulated event samples includes the effect of multiple \(pp\) interactions per bunch crossing (pile-up), as well as the effect on the detector response due to interactions from bunch crossings before or after the one containing the hard interaction. The effect of pile-up was modelled by overlaying the simulated hard-scattering event with inelastic \pp collisions generated with \PYTHIA[8.186]~\cite{Sjostrand:2007gs} using the \textsc{NNPDF2.3lo} PDF set and the A3 set of tuned parameters~\cite{ATL-PHYS-PUB-2016-017}. The simulated events were processed through the same reconstruction algorithms as used for the data.

The measured cross-sections are compared with the general-mass variable-flavour-number scheme (GM-VFNS)~\cite{PhysRevD.71.014018, PhysRevLett.96.012001, PhysRevD.79.094009, Benzke2017, PhysRevD.71.094013} and the fixed-order next-to-leading-logarithm (FONLL)~\cite{Cacciari:1998it,Cacciari2012} predictions.
These calculations provide more reliable predictions than the traditional NLO massive and massless calculations in the intermediate and high transverse momentum range, where the heavy quark mass, $m_{Q}$, is not negligible compared to the transverse momentum of the quark, ${\pt}_Q$.
Most of the setup for GM-VFNS and FONLL is unchanged as described in the previous ATLAS measurements~\cite{BPHY-2013-04}; the only updates are the PDF set and the fragmentation fractions used in the FONLL predictions.

The GM-VFNS prediction aims to combine the massless scheme with the massive scheme.
In the calculation, the charm quark PDF evolves with massless evolution, while the heavy quark mass is retained in the hard-scattering calculation.
With such an approach, the calculation agrees with the massless scheme at ${\pt}_Q \gg m_{Q}$, and with the massive scheme at ${\pt}_Q \approx m_{Q}$.
The GM-VFNS calculation uses the \textsc{CT14nlo}~\cite{Dulat:2015mca} PDF set and the fragmentation functions are taken from the KKKS08 set~\cite{Kneesch:2007ey}.
The prediction is available for both \D and \Ds mesons; the inclusive (prompt and non-prompt) values are provided by the authors themselves.
The uncertainties in the GM-VFNS predictions are dominated by QCD scale uncertainties, namely those of the renormalisation and factorisation scales for initial-state singularities and of the factorisation scale for final-state singularities.

The FONLL calculation consists of three components: the heavy quark production cross-section calculated in perturbative QCD, the non-perturbative heavy-flavour fragmentation, and the decay function describing the heavy hadron decay into leptons.
The principle of FONLL is to expand the massless scheme computation in powers of the strong coupling $\alphas$, and replace a finite number of terms with their massive scheme counterparts.
For instance, the massive and massless scheme predictions are matched exactly up to $O(\alphas^3)$, giving predictions that are reliable for both ${\pt}_Q \approx m_{Q}$ and ${\pt}_Q \gg m_{Q}$ regions.
Currently, the FONLL calculation is available for the \D meson, but not for the \Ds meson~\cite{Cacciari:1998it, Cacciari:2001td, Cacciari2012, Cacciari:2015fta}. The calculation uses the \textsc{NNPDF3.0nlo}~\cite{Ball:2014uwa} PDF set, and the fragmentation fractions $f(b/c \rightarrow D)$ are taken from averaging the LEP measurements~\cite{Gladilin:2014tba}. The fragmentation fractions, updated by the author of Ref. \cite{Gladilin:2014tba} using the newest values of charmed hadron decay branching ratios in Ref. \cite{ParticleDataGroup:2024cfk}, are $f(b \rightarrow \D)=0.217\pm0.011$ and $f(c \rightarrow \D)=0.219 \pm 0.010$.\footnote{\protect\Fragmentation}
The theoretical uncertainties considered include the renormalisation and factorisation scale uncertainties, the pole-mass uncertainties of charm and bottom quarks, the PDF uncertainty, and the fragmentation-fraction uncertainty.

For both GM-VFNS and FONLL, the inclusive cross-sections at \sqrtsthirteen are also compared with the corresponding values at \sqrtseven in the same fiducial volume. In such comparisons, the scale uncertainties at different centre-of-mass energies are assumed to be fully correlated, in order to evaluate the ratios between cross-sections at different centre-of-mass energies.


%

\section{Event reconstruction and selection}
\label{sec:event_selection}
\FloatBarrier

To identify and select the decay $\D/\Ds\to\phi(\mu\mu)\pi^\pm$, a series of selection criteria are defined to enhance the signal-to-background ratio; the selections are summarised in Table~\ref{tab:selection}.
The events are first required to have been accepted by two-muon triggers, which pre-selected muons with certain \pT thresholds and a loose di-muon invariant mass requirement. At least one of the two muons was required to satisfy a \pt threshold of \SI{6}{\GeV}; the threshold for the other muon was either \SI{6}{\GeV} or \SI{11}{\GeV}, depending on the year of data-taking~\cite{TRIG-2018-01}.
Combined ID+MS measurements of track parameters are used to reconstruct muons; they are required to have \pt greater than \SI{6}{\GeV}, to be within the pseudorapidity range of $|\eta|<2.5$, and to satisfy the \textit{Loose} identification working point requirements~\cite{MUON-2018-03}.
Di-muon candidates are also required to contain two selected muons with opposite electric charges $Q_\mu$.
The di-muon invariant mass is required to agree with the $\phi$ meson mass of \SI{1019.5}{\MeV}~\cite{ParticleDataGroup:2024cfk} within a window $\delta m (|\eta|)$ that scales linearly as a function of absolute pseudorapidity $|\eta|$, from \SI{24}{\MeV} at $|\eta|=0$ to \SI{48}{\MeV} at $|\eta|=2.5$.
This requirement retains approximately 90\% of di-muon signal events in MC simulated events.
In addition, one track satisfying the \textit{Loose} identification working point requirements~\cite{ATL-PHYS-PUB-2015-051}, with \pt of at least \SI{1}{\GeV} and within the pseudorapidity range of $|\eta|<2.5$, is required to be reconstructed in the ID\@.

\begin{table}[t]
\begin{center}
\caption{Summary of selection requirements, with items defined in the text.}
\begin{tabular}{lcc}
\hline
\multicolumn{2}{c}{Selection}                                                \\
\hline
Muon objects             &       Two muons satisfying the \textit{Loose}~\cite{MUON-2018-03} working point\\
Track object             &       One track satisfying the \textit{Loose}~\cite{ATL-PHYS-PUB-2015-051} working point\\
Transverse momentum       &       \pTMu $>$ \SI{6}{\GeV}, \pTPi $>$ \SI{1}{\GeV}\\
Opposite charge muons      &      $Q_{\mu_1} \times Q_{\mu_2} = -1$                 \\
Total charge              &      $|Q_{\mu\mu\pi}| = 1$                   \\
Di-muon invariant mass    &  $|m_{\mu\mu} - m_\phi| < \delta m (|\eta|)$     \\
\Lxy{} significance       &       $\text{Sig}(\Lxy{})>3$                   \\
\aZeroxy{} significance   &   $|\text{Sig}(\aZeroxy{})|<4$                 \\
Vertex $p$-value            &      $\log(\pzero)>-0.8$                      \\
Highest vertex $p$-value    &         The vertex with Max($\pzero$) in the event             \\

\hline
\end{tabular}

\label{tab:selection}
\end{center}
\end{table}


All possible di-muon-plus-track ($\mu\mu\pi$) combinations that satisfy the total charge requirement $|Q_{\mu\mu\pi}| = 1$ are used as inputs to $D$ meson candidate secondary vertex (SV) fits, to maximize the number of vertex candidates.
For each candidate, a primary vertex (PV) refit~\cite{ATL-PHYS-PUB-2015-026} is performed after removing the three tracks associated to the candidate, providing an updated PV position.
Due to the \D and \Ds lifetime, the three-particle SV is often separated from the primary vertex. The characteristics of the separation between the SV and the PV are therefore used to reject background.
Two projections of the SV displacement relative to the PV in the transverse plane are used: $\Lxy = |\vec{L}_\mathrm{T}| \cos \theta_\mathrm{xy}$ and $\aZeroxy = |\vec{L}_\mathrm{T}| \sin \theta_{xy}$, where $\vec{L}_\mathrm{T}$ is the vector connecting the PV and the SV in the transverse plane, and $\theta_{xy}$ is the angle between $\vec{L}_\mathrm{T}$ and the transverse momentum $\vec{\pt}$ of the $\mu\mu\pi$ candidate.
Since a large separation in $\Lxy$ distributions is observed between signal and background events, a stringent cut is placed on this variable. For $\aZeroxy$, the background distribution has a tail similar to the non-prompt signal.
To account for detector resolution, the significance of \Lxy and \aZeroxy, $\text{Sig}(\Lxy{})$ and $\text{Sig}(\aZeroxy{})$, are defined as $\Lxy/\sigma_{\Lxy}$ and $\aZeroxy/\sigma_{\aZeroxy}$, where $\sigma_{\Lxy}$ and $\sigma_{\aZeroxy}$ are the uncertainties in \Lxy and \aZeroxy, respectively.
The selections placed on these two variables are $\text{Sig}(\Lxy{})>3$ and $|\text{Sig}(\aZeroxy{})|<4$.

Given the large number of low-\pt tracks originating from hadronic activity, many tracks that are not produced by $D$ meson decays are mistakenly included as candidates. To suppress these combinatorial backgrounds, which are often poorly reconstructed, requirements are placed on the goodness of fits for the SV\@.
The SV $p$-value \pzero is defined as the probability of obtaining a $\chi^2$ value larger than that of the reconstructed SV; only SVs with $\log(\pzero)>-0.8$ (i.e. $\pzero > 0.158$ ) are selected.
Because of the detector coverage and the trigger thresholds, the final candidate is required to have \pt above \SI{12}{\GeV} and to be within the pseudorapidity range of $|\eta|<2.5$. In the following sections, \pt and $\eta$ will refer to the transverse momentum and pseudorapidity of the $\mu\mu\pi$ system unless otherwise specified.
At the end of the selection chain, only the $\mu\mu\pi$ candidate with the highest vertex fit probability $p$-value \pzero is selected per event.


\section{Signal extraction and cross-section measurement}

To extract the signal, the invariant mass of the di-muon-plus-track candidates $m_{\mu\mu\pi}$ is constructed. The yields of the $D$ mesons are then extracted by performing fits to the distributions of $m_{\mu\mu\pi}$ in bins of \pt and $|\eta|$; the model for this fit is described in \cref{sec:analysis_intro}.
Since evaluating the differential cross-section requires reconstruction efficiencies in multiple bins, such values are derived from MC simulated events.
The reconstruction efficiencies are different for $D$ mesons produced promptly and non-promptly, of which the relative contributions can be poorly modelled in MC simulation; therefore, an additional study is performed to constrain the fraction of non-prompt production from data.
This involves extracting the signal yields in bins of pseudo-proper lifetimes and fitting the resulting distributions with templates from MC simulated events corresponding to prompt and non-prompt production; this procedure is described in \cref{sec:bfraction}. Finally, the cross-section calculation is described in \cref{sec:analysis:cross_section}.


\subsection{Invariant mass fit model}
\label{sec:analysis_intro}
Since a fit to the invariant mass $m_{\mu\mu\pi}$ is used to extract the signal yield in bins of \pt, $\eta$, and lifetime, the fit model must be flexible and stable enough to extract yields in subsets of data with different sample sizes and signal-to-background ratios. For the \D and \Ds signals, the fit models $P_{\D}$ and $P_{\Ds}$ are chosen to be non-relativistic Voigtian distributions (Voigt), because of the small number of floating shape parameters (three) and their good performance based on tests in simulated samples.
A Voigtian distribution is the convolution of a Breit--Wigner distribution with a Gaussian (Gauss) distribution~\cite{Voigt1912}.
Since the \D and \Ds mesons have negligible natural widths, the detector resolution is absorbed by the shape parameters of both the Breit--Wigner distribution and the Gaussian distribution.
The background contribution $P_\text{Bkg}$, originating mainly from combinatorics, is extracted using a normalised quadratic exponential distribution. Using RooFit~\cite{RooFit}, the extended unbinned maximum likelihood $\mathcal{L}(m)$ of the combined fit function is implemented as the following:
\begin{align}\label{eq:mass_fit}
\begin{split}
\mathcal{L}(m) &= \frac{\text{e}^{-(S_{\D}+ S_{\Ds} + B)}}{n!}\prod  \left[ S_{\D}  P_{\D}(m) + S_{\Ds}  P_{\Ds}(m) +B  P_\text{Bkg}(m) \right] \times \mathcal{G}(\Delta) ,\\
P_{\D}(m) &= \text{Voigt}(m; m_{\D}, \gamma_{\D}, \sigma_{\D}) ,\\
P_{\Ds}(m) &= \text{Voigt}(m; m_{\Ds}, \gamma_{\Ds}, \sigma_{\Ds}) ,\\
P_\text{Bkg}(m) &= A_\text{norm} \cdot \text{e}^{(c_1 m + c_2 m^2)} ,\\
\mathcal{G}(\Delta) &= \text{Gauss}(\Delta; \mu_{\Delta}, \sigma_{\Delta}) ,\\
\end{split}
\end{align}
where $A_\text{norm}$ is the normalisation factor of the quadratic exponential distribution, and $S_{\D}$, $S_{\Ds}$ and $B$ are the yields of \D mesons, \Ds mesons and background, respectively. The Voigtian distributions of the \D and \Ds mesons are parameterised by the mean parameters $m_{\D}$ and $m_{\Ds}$, while $\gamma_{\D}$ and $\gamma_{\Ds}$ are the widths of the Breit--Wigner distributions and the widths of the Gaussian distributions are $\sigma_{\D}$ and $\sigma_{\Ds}$.
The parameter $c_1$ is the linear rate parameter as in a simple exponential distribution, while the parameter $c_2$ is the quadratic part that is multiplied by $m^2$.
The mass difference $\Delta$ is defined as $m_{\Ds} - m_{\D}$, which is constrained to be close to the world average mass difference $\mu_{\Delta}$~\cite{ParticleDataGroup:2024cfk} of the two $D$ mesons by using an additional Gaussian constraint $\mathcal{G}(\Delta)$ with mean $\mu_{\Delta}$ and width $\sigma_{\Delta}$.

\begin{figure}[thbp]
\centering
\includegraphics[width=0.6\textwidth]{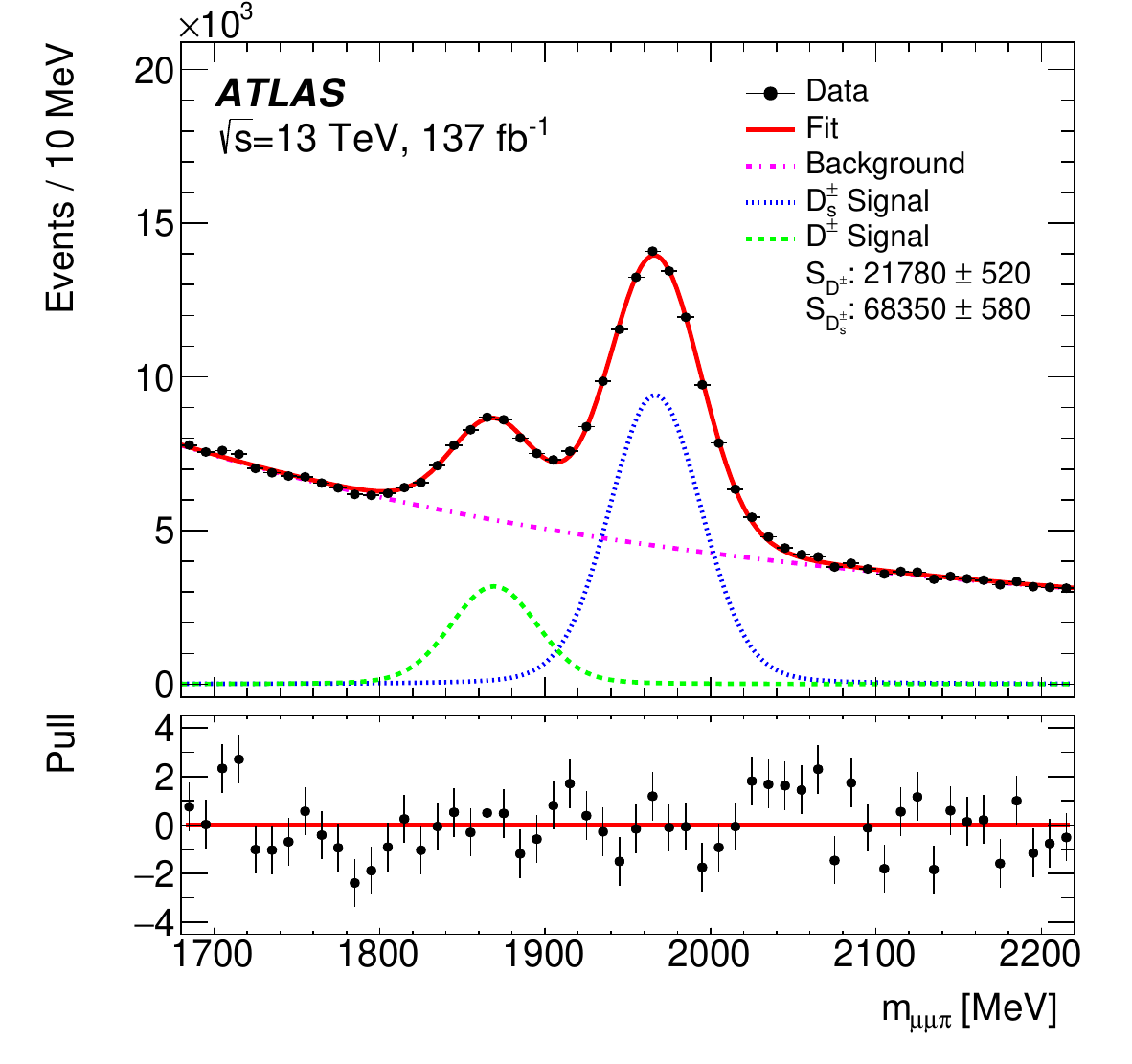}
\caption{Invariant mass distribution of the di-muon-plus-track candidates. The unbinned fit according to Eq.~(\ref{eq:mass_fit}) is shown as a solid line, with the signal components for the \D and \Ds resonance fit presented as dashed and dotted lines, respectively; the background-only contribution is shown as a dash-dotted line. The signal yields extracted are also shown with corresponding statistical uncertainties. The lower panel shows the pull distribution.}
\label{fig:mass_fit}
\end{figure}

In the model, the parameters of interest are the yields $S_{\D}$ and $S_{\Ds}$, while the other parameters that determine the background model and the signal shape are regarded as nuisance parameters.
To improve the stability of the fit model, particularly in bins with low numbers of candidates, fits are performed to the MC simulated events to fix some of the nuisance parameters, including $\gamma_{\D}$ and $\gamma_{\Ds}$.
The unbinned maximum-likelihood fit to the mass spectrum is presented in Figure~\ref{fig:mass_fit}; for each data point, the pull is defined as the difference between the data and the fitted model divided by the statistical uncertainty in the data point.
It can be observed in the figure and in the fits to different kinematic regions in \cref{sec:analysis:cross_section} that this approach achieves compatibility with the observed data while reducing the number of fit parameters.


\subsection{Lifetime fit model}
\label{sec:bfraction}
This section describes the extraction of prompt and non-prompt lifetime templates from the MC simulation, together with a template fit to the data to estimate the contribution from non-prompt processes. To quantify the proportion of non-prompt processes, the non-prompt fraction $f_\mathrm{NP}$ is defined as the yield of non-prompt signal divided by the total signal yield.
The two production mechanisms mainly differ in terms of the average decay distance from the PV\@.
From the reconstructed vertex position, the pseudo-proper-lifetime (hereinafter referred to as lifetime) $\tau$ of the $\mu\mu\pi$ candidates is calculated as:
\begin{equation}
\tau = \frac{m_{\mu\mu\pi}  \Lxy}{\pt}.
\end{equation}

Because of the relatively long lifetime of \D (\Ds) mesons, they can travel for approximately \SI{1}{\ps} (\SI{0.5}{\ps}) before they decay into a $\phi$ meson and a pion. The prompt contribution therefore includes an exponential distribution (Exp) to describe the physical exponential decay of the particle, which is convolved with a Gaussian distribution (Gauss) and an error function (Erf) to account for the detector resolution and the turn-on effect at low lifetime. The turn-on effect describes the transition in efficiency from zero to a constant at low lifetime, which is mainly due to the requirement on the significance of \Lxy.

For the non-prompt production mode, the $B$ meson, which has a typical lifetime of \SI{1.5}{\ps}, can travel a few millimeters before it decays into a \D/\Ds meson. The non-prompt contribution therefore includes one exponential distribution to account for the cascade decay of the $B$ meson, and another exponential distribution to account for the lifetime of the \D/\Ds mesons. As for  the prompt contribution, the distribution is also convolved with a Gaussian distribution and an error function to account for detector effects.

The templates for the non-prompt contribution $P_{\bb}(\tau)$ and the template for the prompt contribution $P_{\cc}(\tau)$ are defined as follows:
\begin{align}
\begin{split}
P_{\bb}(\tau)  &=  \text{Exp}(\tau; \tau^{\bb}_{D}) \ast \text{Exp}(\tau; \tau^{\bb}_{B}) \ast \text{Gauss}(\tau; \mu^{\bb}, \sigma^{\bb}_\text{res} ) \ast \text{Erf}( \tau; \tau^{\bb}_\text{turn-on}, \beta^{\bb}) ,\\
P_{\cc}(\tau)  &=  \text{Exp}(\tau; \tau^{\cc}_{D}) \ast \text{Gauss}(\tau; \mu^{\cc}, \sigma^{\cc}_\text{res} ) \ast \text{Erf}( \tau; \tau^{\cc}_\text{turn-on}, \beta^{\cc}) .\\
\end{split}
\end{align}
Here the asterisk sign $\ast$ indicates the linear convolution between the functions, $\tau$ is the lifetime of the secondary vertex, $\tau^{\bb}_{D}$ and $\tau^{\cc}_{D}$ are the lifetime components of the \D/\Ds meson, $\tau^{\bb}_{B}$ is the lifetime component of the $B$ meson for the non-prompt decay, $\mu^{\bb}$ and $\mu^{\cc}$ represent a constant shift in lifetime due to detector bias, $\sigma^{\bb}_\text{res}$ and $\sigma^{\cc}_\text{res}$ describe the smearing due to detector resolution, $\tau^{\bb}_\text{turn-on}$ and $\tau^{\cc}_\text{turn-on}$ represent a constant shift in lifetime due to the turn-on effect, and $\beta^{\bb}$ and $\beta^{\cc}$ are multiplicative factors applied to the lifetime that determine the sharpness of the turn-on effect. These functions are fitted to the prompt and non-prompt MC simulated events separately to extract the templates, and the parameters are fixed before extracting the non-prompt fraction from data.

Before performing a lifetime fit to the data using the MC templates, the signal distribution must be extracted from data by removing the background contribution.
Invariant mass fits (as described in \cref{sec:analysis_intro}) are performed in bins of the lifetime, $\tau$, allowing the signal yield $N_i$ and its uncertainty $\sigma_i$ to be extracted in each lifetime interval $i$ for the data.
A minimum $\chi ^ 2$ fit is performed on the extracted signal distribution using the prompt and non-prompt lifetime templates.
Since the parameters of the individual MC templates are fixed, the only fit parameter is the non-prompt fraction $f_\mathrm{NP}$.
With such an approach, the fits can extract yields up to lifetimes of \SI{5.0}{\ps} for the \Ds meson and \SI{8.0}{\ps} for the \D meson; beyond this, there are insufficient data to perform the fits.
Because the signal yields are extracted in finite bins of lifetime, the sharply rising slope at low lifetime cannot be reliably extracted; therefore, the region with low lifetime is excluded from the fit.
For the \D meson, the region with lifetime less than \SI{1.0}{\ps} is excluded; for the \Ds meson, the region with lifetime less than \SI{0.5}{\ps} is excluded.

\begin{figure}[thbp]
\centering
\subfloat[][]{\includegraphics[width=0.5\textwidth]{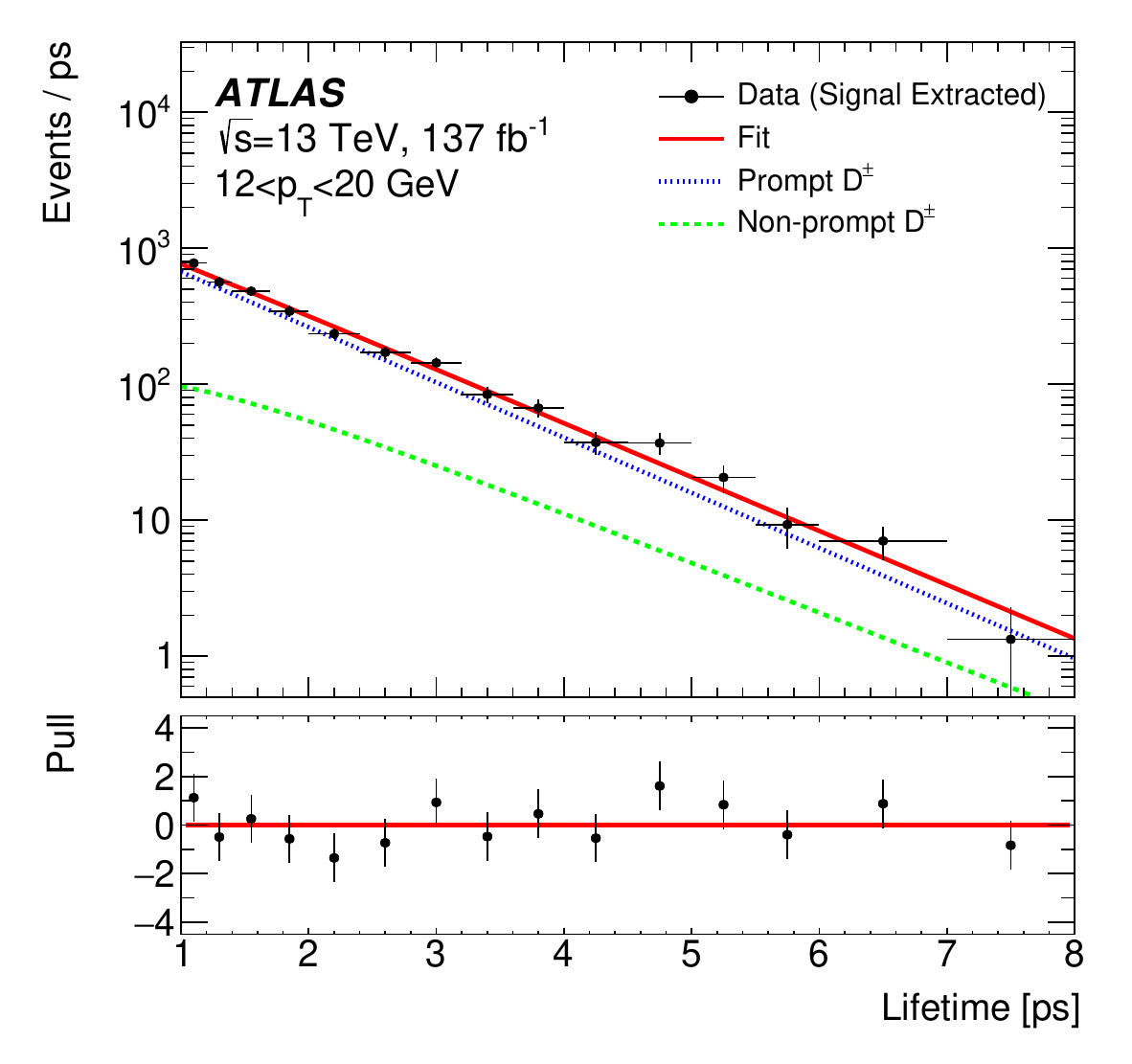}}
\subfloat[][]{\includegraphics[width=0.5\textwidth]{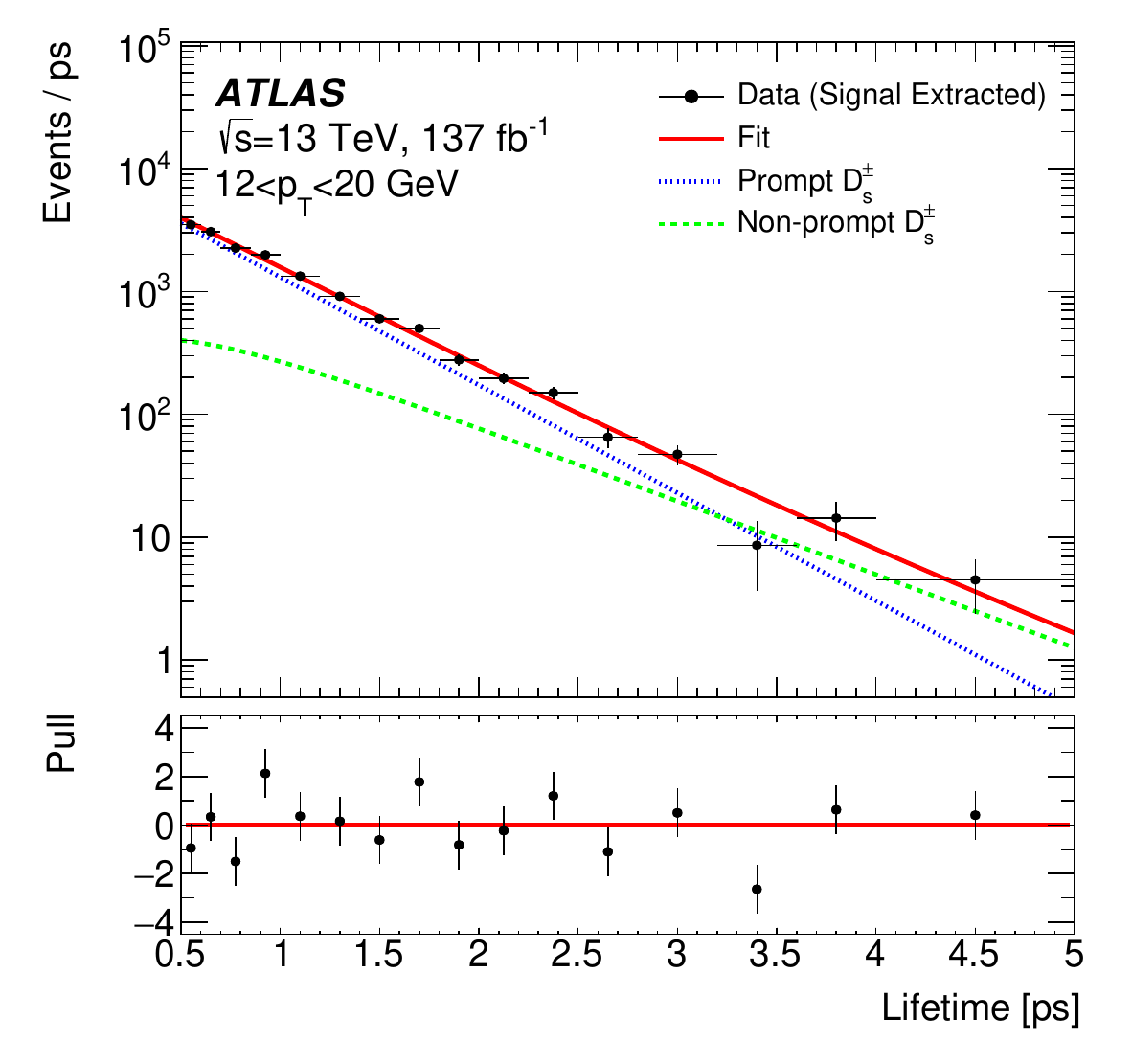}}
\caption{Template fits to the lifetime distribution extracted from data for (a) \D and (b) \Ds mesons in the range of 12 $< \pt <$ 20 \GeV. The prompt and non-prompt contributions are presented as dotted and dashed lines, respectively; the combined fits are shown as solid lines. The bin sizes are variable, and the data points are drawn at the bin centres. The lower panels show the pull distributions.}
\label{fig:lifetime:fit_bfrac_dpm}
\end{figure}

The procedure is applied separately for \D and \Ds mesons across three bins of \pt; in each bin of \pt, the individual templates and the combined models are all fitted separately.
The bins are defined to capture the falling shape of $f_\mathrm{NP}$, while ensuring enough data are present in each bin. The bin boundaries in \pt are [12, 20, 30, 100] \GeV.
\cref{fig:lifetime:fit_bfrac_dpm} shows the extracted signal from data and the fits with the combined model for the first \pt region.
The $\chi^2$ divided by the number of degrees of freedom of the fits ranges from 0.6 to 1.5, showing a good agreement between the model and the data.

\begin{figure}[thbp]
\centering
\subfloat[][]{\includegraphics[width=0.5\textwidth]{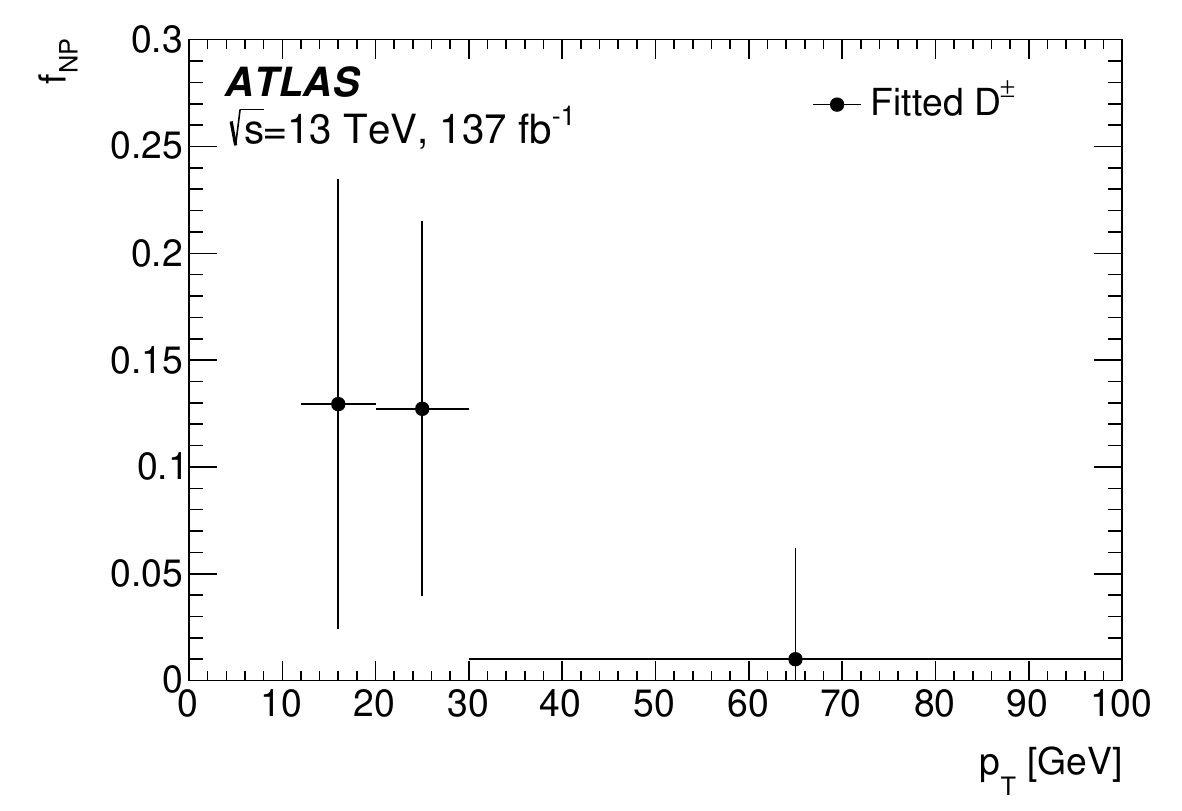}}
\subfloat[][]{\includegraphics[width=0.5\textwidth]{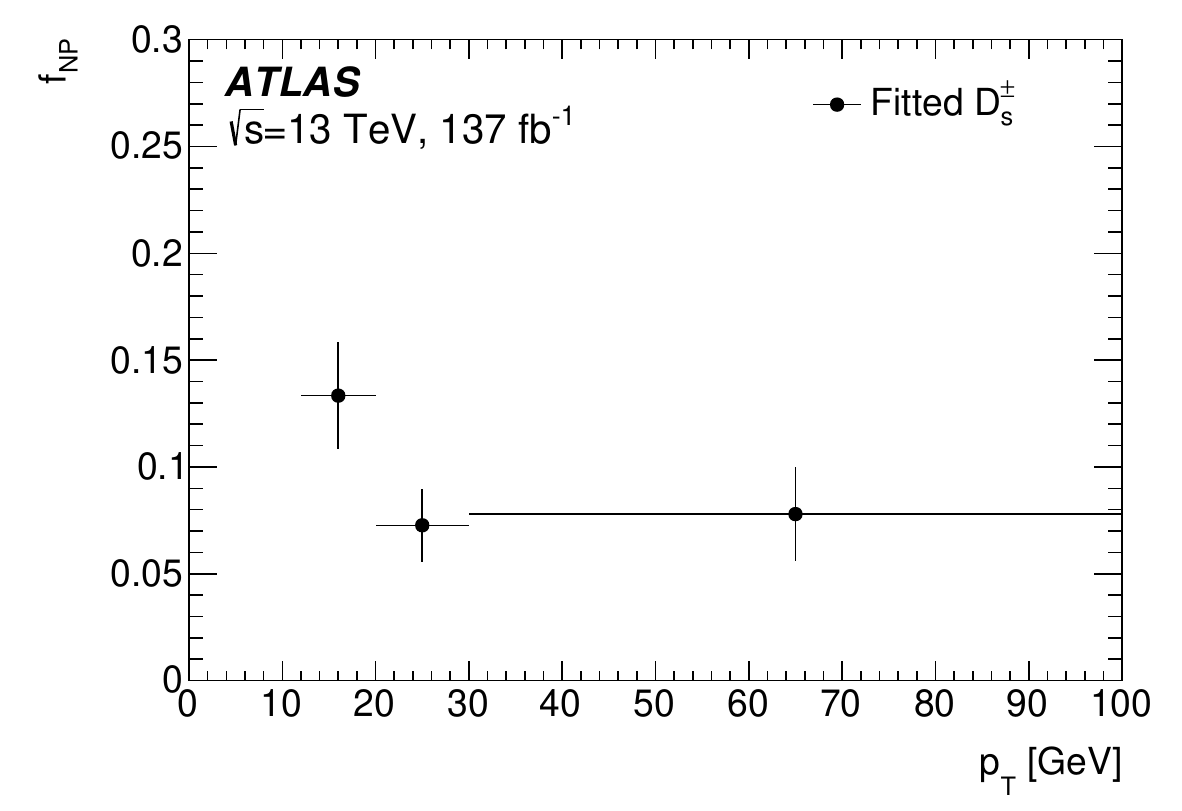}}
\caption{Estimated non-prompt fraction of (a) \D and (b) \Ds meson production as a function of \pt, in the fiducial volume defined by $12<\pt<100$~\SI{}{\GeV} and $|\eta|<2.5$. Only the statistical uncertainties are shown.}
\label{fig:lifetime:f_bb}
\end{figure}

Figure~\ref{fig:lifetime:f_bb} shows the extracted non-prompt fraction and the corresponding statistical uncertainties for both \D and \Ds in bins of \pt.
The statistical uncertainties for \D mesons are larger than those for \Ds mesons.
This is due to the lower yield and the longer lifetime of the \D meson, which reduces the discriminating power of the fit because of the increased similarity between the prompt and non-prompt shapes.

The non-prompt fraction is then used to construct the weights for the MC simulated events, such that the non-prompt contribution in data is matched in the MC simulated events.
The uncertainties in the non-prompt fractions are relatively large; therefore, no interpretation or comparison is made with theoretical predictions.
However, the extracted uncertainties in the non-prompt fractions allow the systematic uncertainties for the two production mechanisms to be evaluated.


\subsection{Cross-section measurement}
\label{sec:analysis:cross_section}

\begin{figure}[b]
\centering
\subfloat[]{\includegraphics[width=0.5\textwidth]{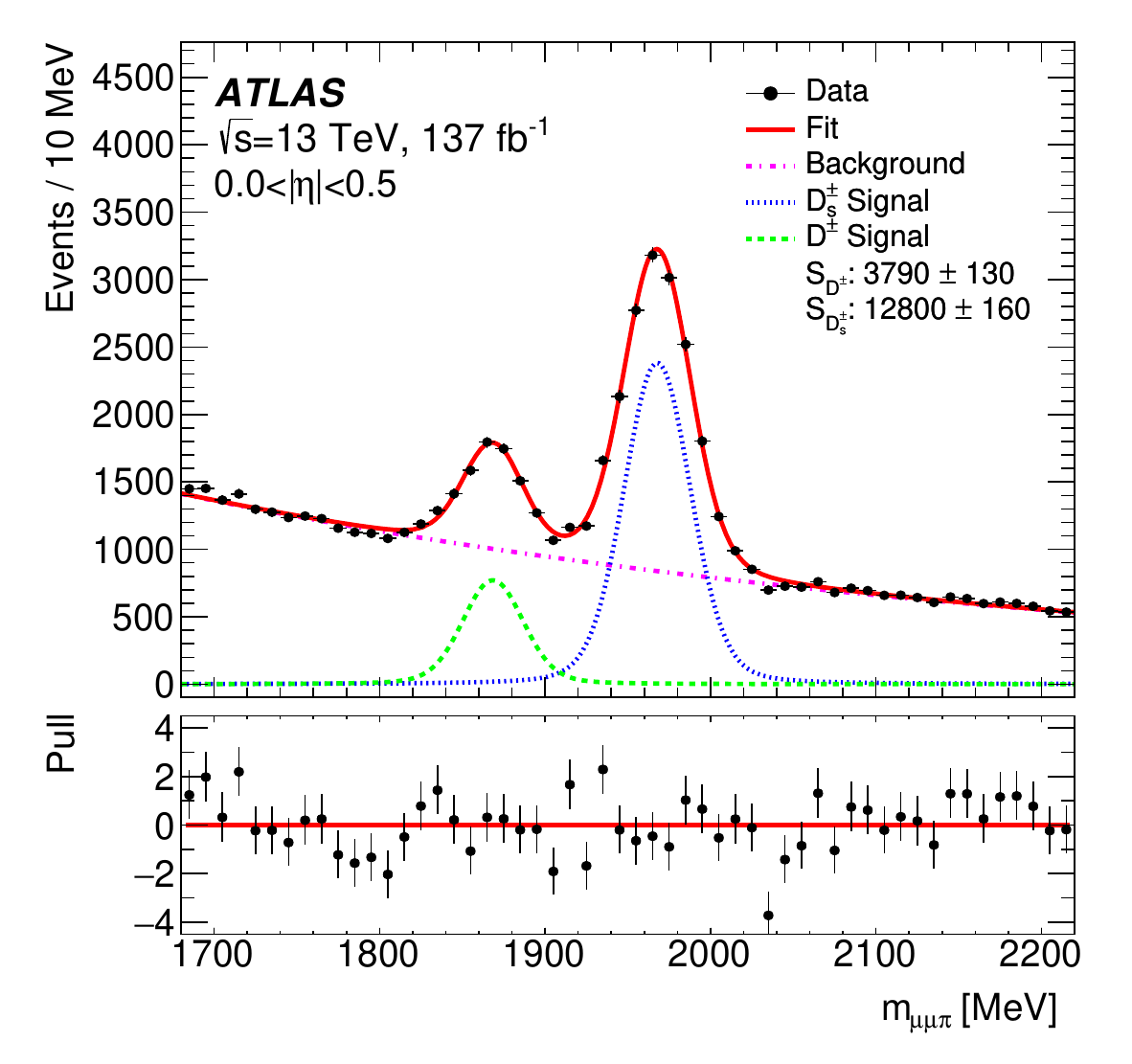}}
\subfloat[]{\includegraphics[width=0.5\textwidth]{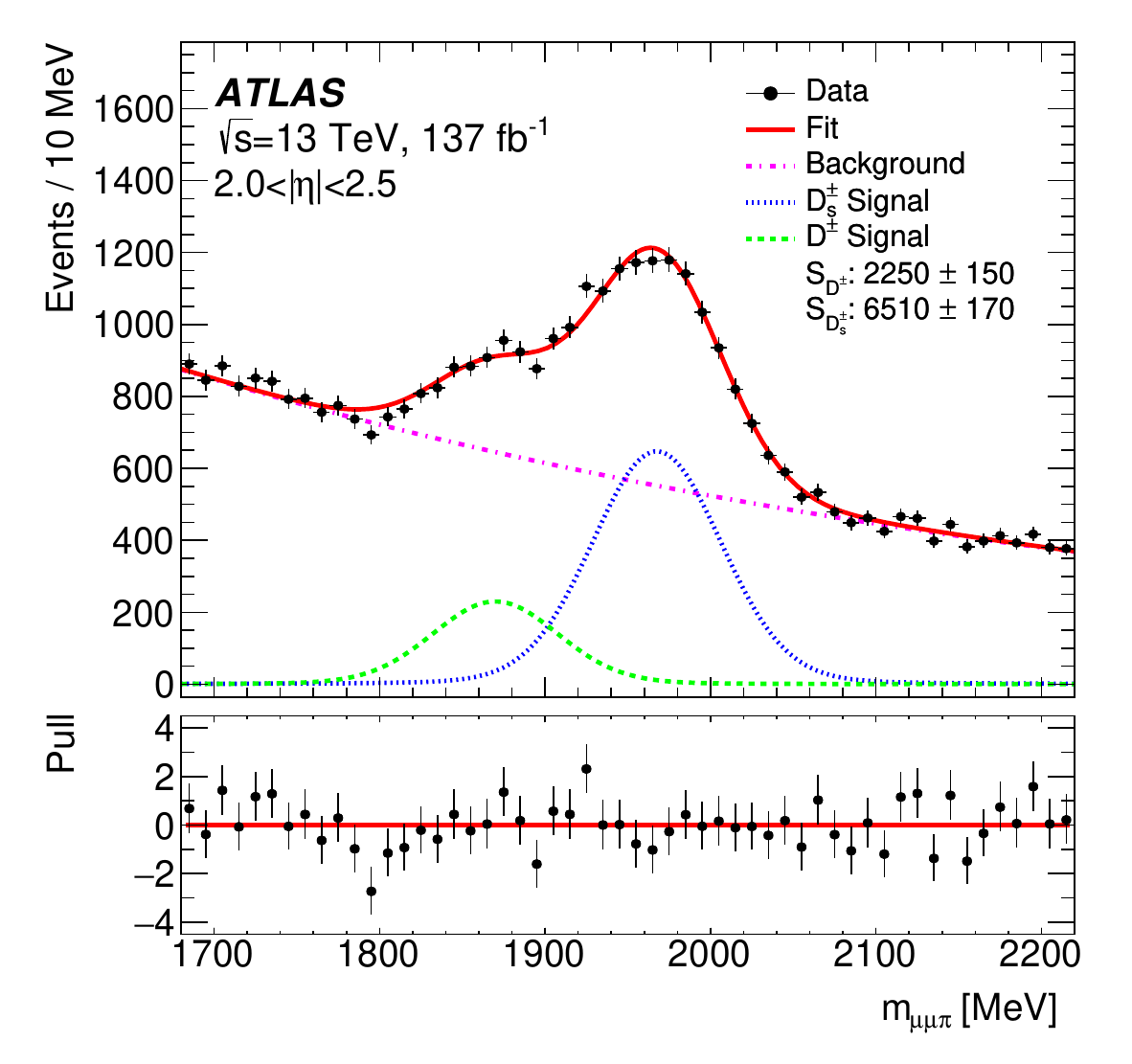}}
\caption{Examples of fits to the distribution of invariant mass $m_{\mu\mu\pi}$ in regions of (a) $0.0 < |\eta| < 0.5$ and (b) $2.0 < |\eta| < 2.5$. The unbinned fits according to Eq.~(\ref{eq:mass_fit}) are shown as solid lines, with the signal components for the \D and \Ds resonance fit presented as dashed and dotted lines, respectively; the background-only contributions are shown as dash-dotted lines. The signal yields extracted are also shown with corresponding statistical uncertainties. The lower panels show the pull distributions.}
\label{fig:fits_eta}
\end{figure}

\begin{figure}[tbhp]
\centering
\subfloat[]{\includegraphics[width=0.5\textwidth]{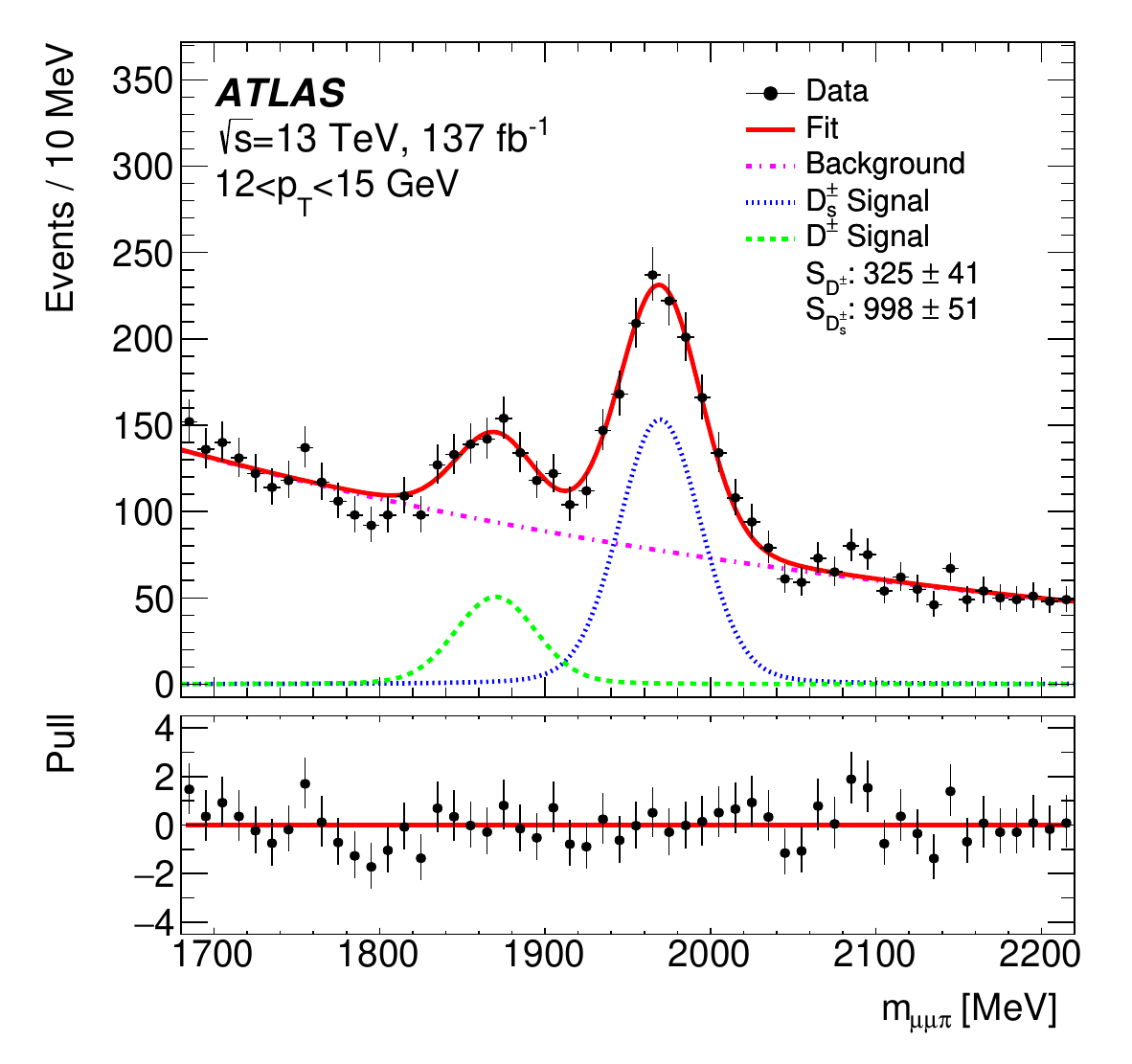}}
\subfloat[]{\includegraphics[width=0.5\textwidth]{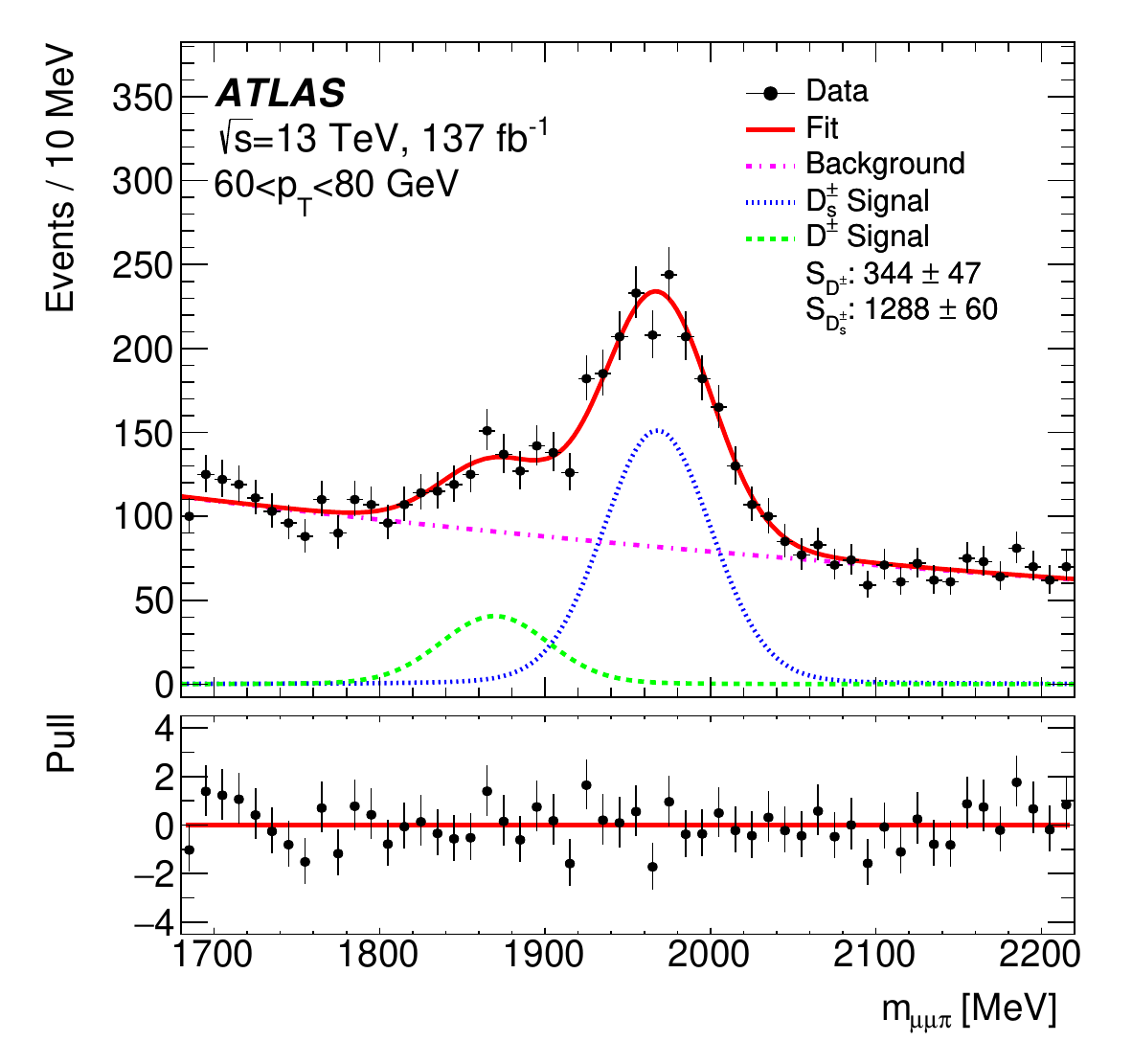}}
\caption{Examples of fits to the distribution of invariant mass $m_{\mu\mu\pi}$ in regions of (a) $12 < \pt < 15$~\GeV and (b) $60 < \pt < 80$~\GeV. The unbinned fits according to Eq.~(\ref{eq:mass_fit}) are shown as solid lines, with the signal components for the \D and \Ds resonance fit presented as dashed and dotted lines, respectively; the background-only contributions are shown as dash-dotted lines. The signal yields extracted are also shown with corresponding statistical uncertainties. The lower panels show the pull distributions.}
\label{fig:fits_pt}
\end{figure}

The differential cross-section is measured in the fiducial volume defined by $\SI{12}{} < \pt < \SI{100}{\GeV}$ and $|\eta| < 2.5$ for \D and \Ds mesons, in nine bins of \pt and five bins in $|\eta|$.
To obtain the differential cross-section in a given bin, the yields of \D and \Ds mesons in the bin are obtained from a fit to the triplet mass $m_{\mu\mu\pi}$, as described in \cref{sec:analysis_intro}. \cref{fig:fits_eta,fig:fits_pt} show the examples of the invariant mass fits in two regions of $|\eta|$ and \pt, respectively.

For each bin in \pt (denoted by $i$) and each bin in $|\eta|$ (denoted by $j$), the yield obtained is then divided by the overall efficiency, bin width, branching ratio $\mathcal{B}$, and the integrated luminosity to evaluate the differential production cross-section.
The differential cross-sections in \pt and $|\eta|$ are given by the following equations:
\begin{equation}
\begin{split}
\left.\frac{\mathrm{d}\sigma}{\mathrm{d}{p_\text{T}}}\right\vert_{i} &= \frac{S^i_{\D/\Ds}}{\int\mathcal{L}\text{d}t \times C^i \times \mathcal{B}(\D/\Ds\rightarrow\phi(\mu\mu)\pi^\pm) \times \Delta^i\pt } ,\\
\left.\frac{\mathrm{d}\sigma}{\mathrm{d}{|\eta|}}\right\vert_{j} &= \frac{S^j_{\D/\Ds}}{\int\mathcal{L}\text{d}t \times C^j \times \mathcal{B}(\D/\Ds\rightarrow\phi(\mu\mu)\pi^\pm) \times \Delta^j|\eta| } ,\\
\end{split}
\label{eq:cross_section}
\end{equation}
where $S^i_{\D/\Ds}$ and $S^j_{\D/\Ds}$ are the yields of \D and \Ds signals extracted in bins of \pt and $|\eta|$, $\Delta^i\pt$ is the bin width in \pt, and $\Delta^j|\eta|$ is the bin width in $|\eta|$. The integrated luminosity, $\int\mathcal{L}\text{d}t$, is \lumi ~\cite{DAPR-2021-01}.
The correction factors $C^i$ and $C^j$ account for the reconstruction and selection efficiency and acceptance, which are derived bin-by-bin in \pt and $|\eta|$, respectively, using MC simulated events.

The branching ratio of the \Ds decay chain is given by:
\begin{equation}
\mathcal{B}(\Ds\rightarrow\phi(\mu\mu)\pi^\pm) = \frac{\mathcal{B}(\Ds\rightarrow\phi(K^+K^- )\pi^\pm)}{\mathcal{B}(\phi\rightarrow K^+K^-)} \times \mathcal{B}(\phi\rightarrow\mu\mu) ,
\end{equation}
as the uncertainties of the world average branching ratios of the $\Ds\rightarrow\phi(K^+K^-)\pi^\pm$ and $\phi\rightarrow K^+K^-$ processes combined are better than the branching ratio of $\Ds\rightarrow\phi\pi^\pm$ ~\cite{ParticleDataGroup:2024cfk}. For the \D meson, the branching ratio is taken as the product of the branching ratios of the $\D\rightarrow\phi\pi^\pm$ and $\phi\rightarrow\mu\mu$ processes ~\cite{ParticleDataGroup:2024cfk}.

The cross-section in the fiducial volume is determined by summing over the bins $i$ of the differential cross-section as a function of \pt, according to:
\begin{equation}
\sigma_\text{fiducial} = \sum_i \left.\frac{\mathrm{d}\sigma}{\mathrm{d}\pt}\right\vert_{i} \Delta^i\pt .
\end{equation}
Summing over the differential cross-section in bins of $|\eta|$ is found to give a consistent result when compared to summing over \pt.

To facilitate the use of the result as a normalisation for other studies and as a comparison with previous measurements, different fiducial volumes are considered.
The lower \pt boundary of \SI{12}{\GeV} corresponds to the lowest \pt reach; the \pt boundaries of 15 and \SI{20}{\GeV} are also considered.
The \pt boundary of \SI{15}{\GeV} provides a \pt range that is above the trigger efficiency plateau to reduce the muon and trigger uncertainties; the \pt boundary of \SI{20}{\GeV} provides a \pt range compatible with the previous ATLAS measurements~\cite{BPHY-2013-04}.
%



\section{Systematic uncertainties}

The uncertainties in the signal yields, acceptance correction factors, integrated luminosity, and the decay branching ratios are propagated linearly to the differential and thus to the integrated fiducial cross-sections according to Eq.~(\ref{eq:cross_section}).
The systematic uncertainties can be categorised into detector effects, production modelling, branching ratio, MC simulated data sample size and signal extraction.

The systematic uncertainties due to detector effects are as follows:
\setlist[description]{leftmargin=0.5cm,labelindent=0.5cm}
\begin{description}
\item[Muon:] The uncertainties on the muon reconstruction were evaluated in dedicated studies using $J/\psi\rightarrow\mu\mu$ and $Z\rightarrow\mu\mu$ events~\cite{MUON-2018-03, MUON-2022-01}. The muon reconstruction and identification efficiency uncertainties~\cite{MUON-2018-03} affect the acceptance correction factors in Eq.~(\ref{eq:cross_section}), while those on muon momentum calibration~\cite{MUON-2022-01} can also cause bin migrations in the differential measurements.

\item[Tracking:] The uncertainty in the track reconstruction efficiency arises primarily from the limited knowledge of the ID material description used in MC simulation. Uncertainties in track momentum and impact parameter calibration are also accounted for in vertex fitting and selection~\cite{PERF-2015-08}.

\item[Pile-up:] Pile-up refers to the number of simultaneous \pp collisions in the same event. Weights are applied to MC simulation to make the pile-up distribution match that in data and are varied according to its uncertainty.

\item[Trigger:] The trigger efficiencies are measured in data using $J/\psi\rightarrow\mu\mu$ events following the procedure described in Ref.~\cite{BPHY-2011-06}. The MC simulated samples are corrected to reproduce the measured efficiencies, and the uncertainty of the measurement is accounted for as a systematic uncertainty.

\item[Luminosity:] The uncertainty in the integrated luminosity of the combined data sample from 2016, 2017, and 2018 is 0.84\%, using the same methodology as in Ref.~\cite{DAPR-2021-01}, and obtained using the LUCID-2 detector~\cite{LUCID2} for the primary luminosity measurements, complemented by measurements using the ID and calorimeters.

\end{description}
The systematic uncertainties due to the modelling of \D and \Ds meson production are as follows:
\setlist[description]{leftmargin=0.5cm,labelindent=0.5cm}
\begin{description}
\item[Non-prompt:] To correct for the MC modelling of the non-prompt \D and \Ds meson production fractions, the procedure described in \cref{sec:bfraction} is used. The uncertainties on the non-prompt fractions are propagated to the yields of the simulated samples, and further propagated to the acceptance correction factors.

\item[Kinematics:] The MC simulated \pt spectra of \D and \Ds meson production are corrected to match the distribution extracted from data. The statistical uncertainties in the data spectra are propagated to the acceptance correction factors using this procedure. The uncertainties due to the $|\eta|$ spectra modelling are found to be negligible, while those of the \pt spectra have a small effect when propagated to the differential cross-section measurements in $|\eta|$.

\end{description}

The remaining systematic uncertainties are as follows:
\setlist[description]{leftmargin=0.5cm,labelindent=0.5cm}
\begin{description}
\item[Branching ratio (BR):] The uncertainties in the branching ratios are taken from Ref.~\cite{ParticleDataGroup:2024cfk}. For the $\D$ meson, the total uncertainty in the decay chain is 7.2\%, while that of the \Ds meson is 7.3\%.
\item[MC size:] The statistical uncertainty due to limited size of MC simulated signal samples is propagated to the measurement results via the acceptance correction factors.
\item[Fit model (signal and background):] The fit model systematic uncertainties can be separated into contributions from the signal model and the background model. By varying the models, the difference in yield is taken as a systematic uncertainty. MC pseudo data are generated according to the distribution obtained from data, according to Poisson fluctuations; fits with alternative models are then applied to the pseudo data. For the signal model, the Bukin model~\cite{Bukin:2007zha} and triple Gaussian models are chosen as alternatives. For the background model, a simple exponential model is taken as the alternative. By repeatedly fitting the nominal models and the alternative models to the pseudo data, the distributions of the yield estimators are obtained. The mean yield differences between the nominal model and the alternative models are taken as the systematic uncertainties. The background model systematic uncertainty is also checked to cover the effects of possible contributions from partially reconstructed $D$ meson decays, such as $\D/\Ds \to \phi \pi^{\pm} \pi^{0}$ decays. Because of the limited amount of data, these uncertainties are evaluated in a single bin above $\pt = \SI{40}{\GeV}$.
\end{description}

\begin{figure}[tbhp]
\centering
\subfloat[][]{\includegraphics[width=0.5\textwidth]{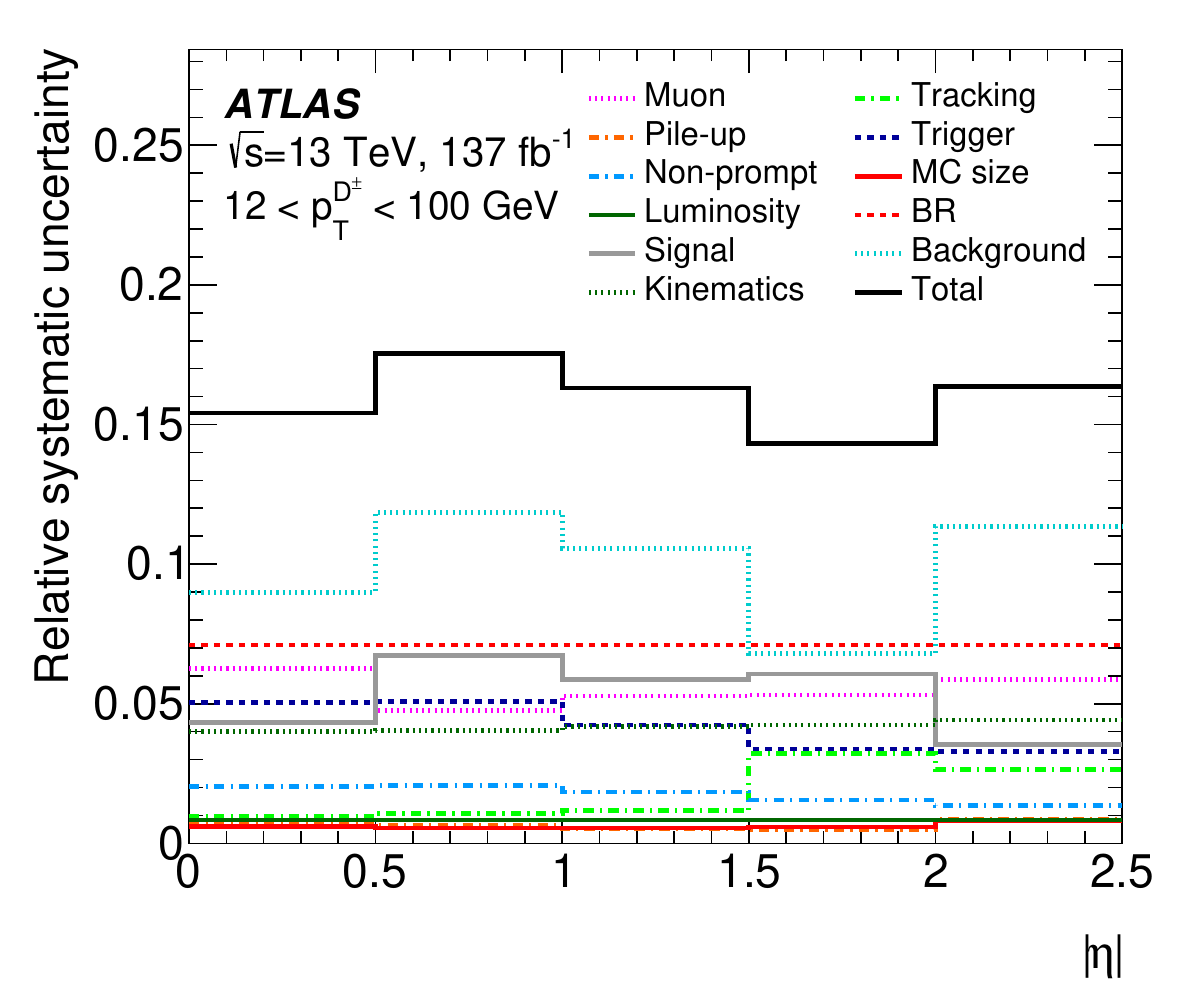}}
\subfloat[][]{\includegraphics[width=0.5\textwidth]{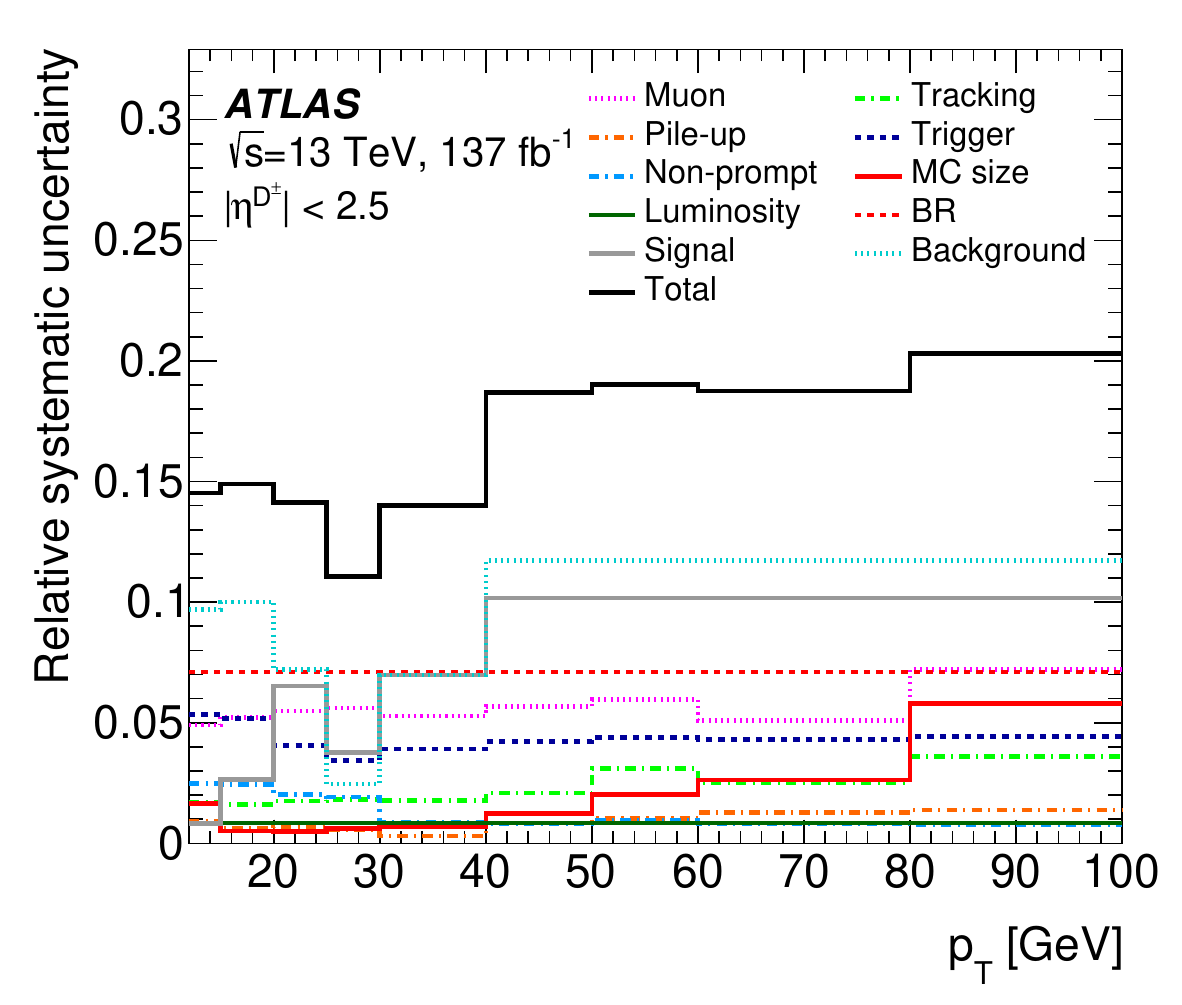}}
\caption{Relative systematic uncertainty profile of the differential cross-sections for \D mesons, in bins of (a) $|\eta|$ and (b) $\pt$. The systematic uncertainties shown are also combined in quadrature (upper solid line).}
\label{fig:systs_dpm}
\end{figure}

\begin{figure}[tbhp]
\centering
\subfloat[][]{\includegraphics[width=0.5\textwidth]{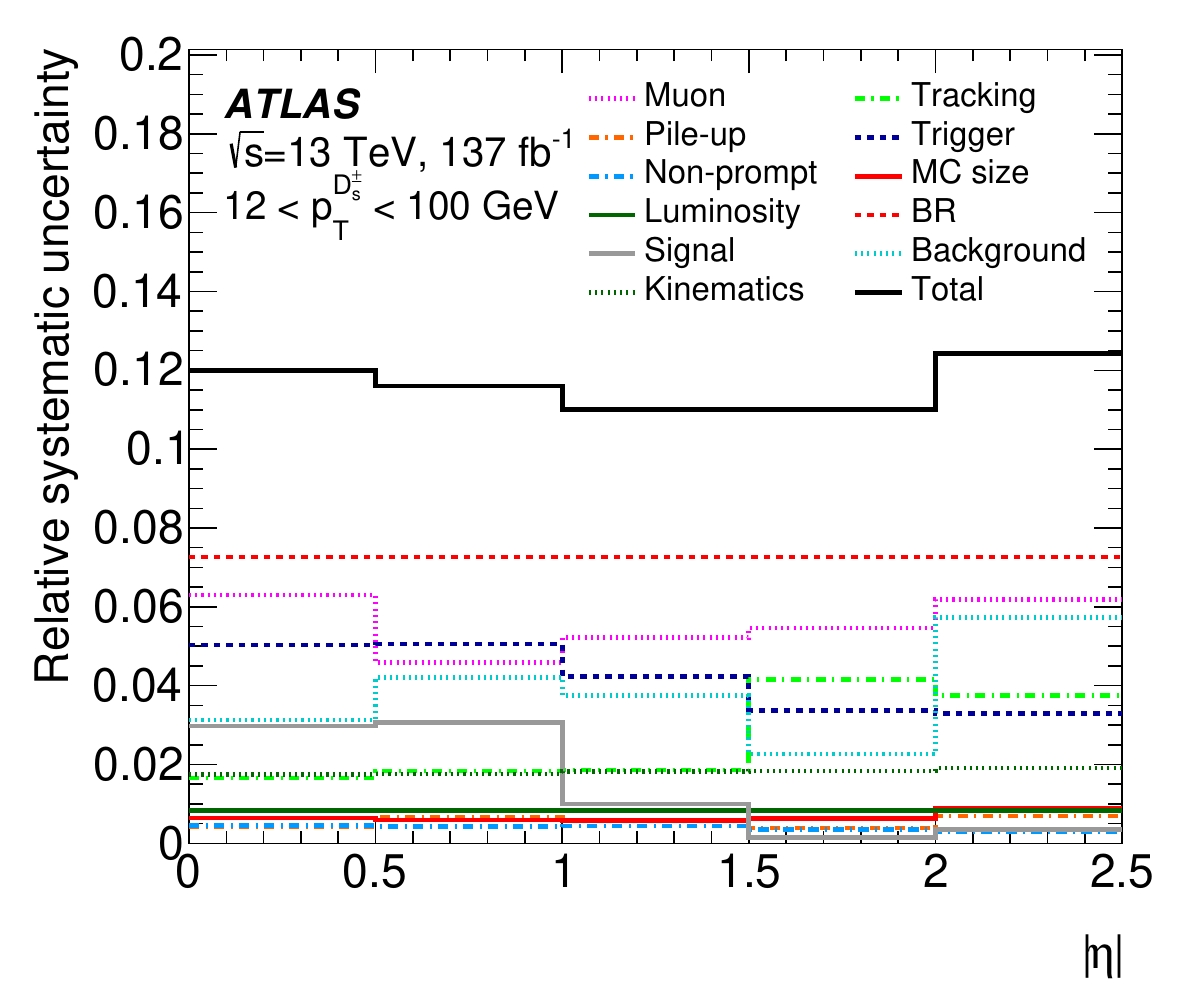}}
\subfloat[][]{\includegraphics[width=0.5\textwidth]{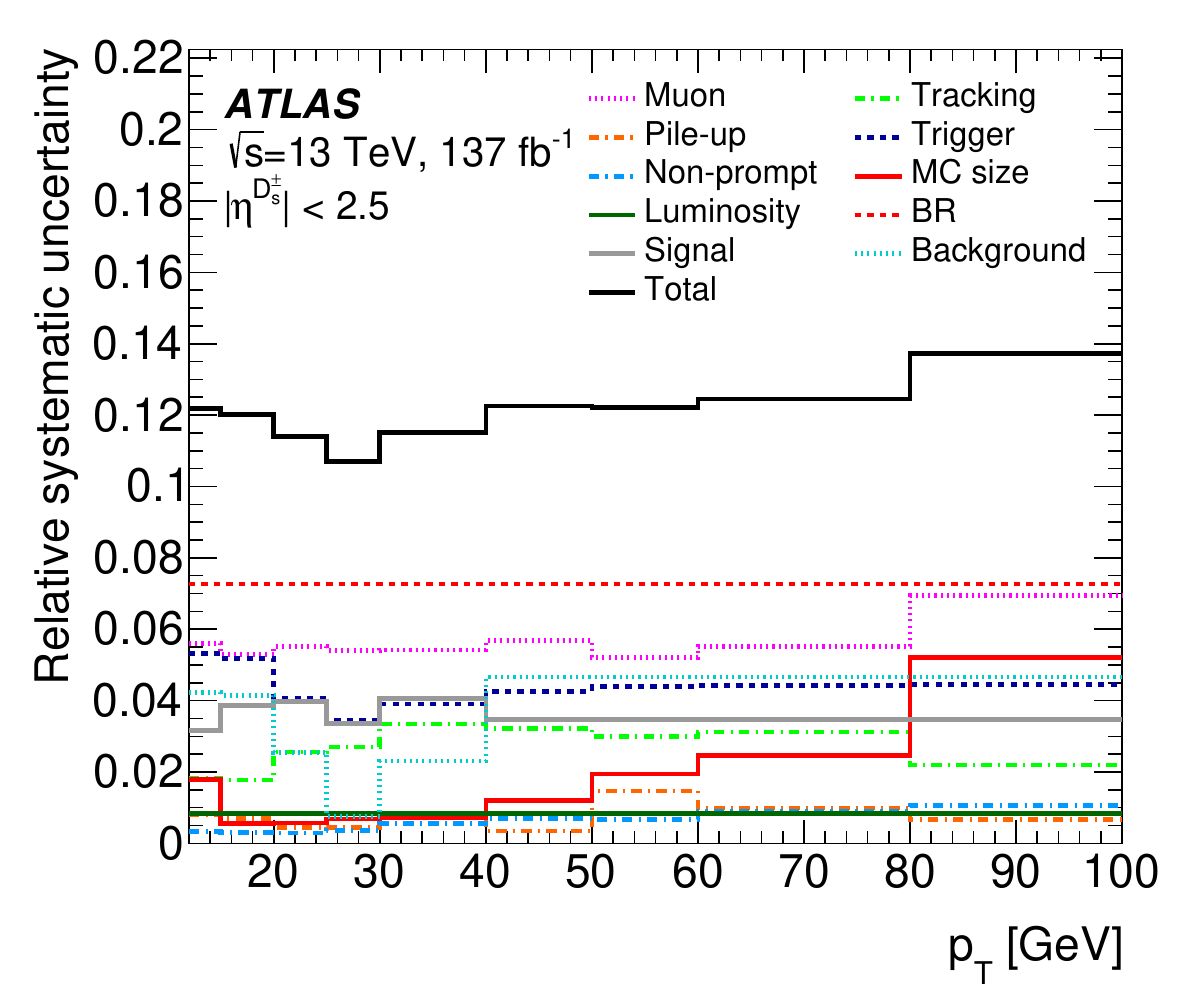}}
\caption{Relative systematic uncertainty profile of the differential cross-sections for \Ds mesons, in bins of (a) $|\eta|$ and (b) $\pt$. The systematic uncertainties shown are also combined in quadrature (upper solid line).}
\label{fig:systs_ds}
\end{figure}

The uncertainties above are evaluated in bins of \pt and $|\eta|$, except for the uncertainties in the luminosity and branching ratios.
\cref{fig:systs_dpm,fig:systs_ds} show the systematic uncertainty profiles of \D and \Ds mesons in bins of \pt and $|\eta|$.

For the \Ds measurement, the largest systematic uncertainty contributions arise from the decay branching ratios, followed by the trigger uncertainty and background modelling uncertainty.  For the \D cross-section, the background modelling uncertainty dominates in most of the phase space. The combined systematic uncertainty is mainly in the range of 10\%--14\% for the \Ds and 15\%--20\% for the \D due to the latter being more sensitive to the choice of background model.

\FloatBarrier

\section{Results}
\FloatBarrier

The values of the measured differential cross-sections and the theory predictions are shown in \Cref{tab:XScontentdpmetaoriginal,tab:XScontentdpmptoriginal,tab:XScontentdsetaoriginal,tab:XScontentdsptoriginal} for \D and \Ds mesons; they are also shown in  Figure~\ref{fig:dXS_dq}.
The fiducial cross-section is summarised in Table~\ref{tab:xsec:Ds}, including the theory predictions available.

\begin{DIFnomarkup}

\begin{DIFnomarkup}

\begin{table}[thbp]
\renewcommand{\arraystretch}{1.5}
\small
\centering
\caption{The measured differential cross-sections and the predictions from GM-VFNS and FONLL calculations for the \D meson in bins of $|\eta|$ for $\SI{12}{} < \pt < \SI{100}{\GeV}$. The statistical, systematic (excluding branching ratio) and branching ratio uncertainties are shown separately for data, while the total theoretical uncertainties are shown for GM-VFNS and FONLL.}
\begin{tabular}{c @{\hspace{0.5cm}}@{}c@{}@{ }d{2.2}@{${}\pm{}$}d{1.2}@{${}\pm{}$}d{1.2}@{${}\pm{}$}d{1.2}@{ }r r r}
\toprule
& \multicolumn{8}{c}{Differential cross-section [\mub]} \\[0.1cm]
Range & \multicolumn{6}{c}{Data} & \multicolumn{1}{c}{GM-VFNS}  & \multicolumn{1}{c}{FONLL} \\
& \multicolumn{6}{c}{$\frac{\mathrm{d}\sigma}{\mathrm{d}|\eta|} \pm \delta_\text{stat} \pm \delta_\text{syst} \pm \delta_\text{BR}$} & \multicolumn{1}{c}{$\frac{\mathrm{d}\sigma}{\mathrm{d}|\eta|} \pm \delta_\text{theory}$ }& \multicolumn{1}{c}{$\frac{\mathrm{d}\sigma}{\mathrm{d}|\eta|} \pm \delta_\text{theory}$}\\
\midrule
$0.0 < |\eta| < 0.5$ & & 5.15 & 0.18 & 0.71 & 0.37 & & $6.4\numpmerr{+1.3}{-1.1}$ & $4.7\numpmerr{+1.2}{-0.9}$ \\
$0.5 < |\eta| < 1.0$ & & 5.84 & 0.19 & 0.93 & 0.41 & & $6.2\numpmerr{+1.3}{-1.0}$ & $4.5\numpmerr{+1.2}{-0.9}$ \\
$1.0 < |\eta| < 1.5$ & & 4.96 & 0.22 & 0.73 & 0.35 & & $5.8\numpmerr{+1.2}{-1.0}$ & $4.2\numpmerr{+1.1}{-0.8}$ \\
$1.5 < |\eta| < 2.0$ & & 3.62 & 0.18 & 0.43 & 0.26 & & $5.3\numpmerr{+1.1}{-0.9}$ & $3.8\numpmerr{+1.0}{-0.7}$ \\
$2.0 < |\eta| < 2.5$ & & 3.33 & 0.23 & 0.49 & 0.24 & & $4.5\numpmerr{+0.9}{-0.7}$ & $3.2\numpmerr{+0.9}{-0.6}$ \\
\bottomrule
\end{tabular}
\label{tab:XScontentdpmetaoriginal}
\end{table}

\end{DIFnomarkup}


\begin{DIFnomarkup}
\begin{table}[thbp]
\renewcommand{\arraystretch}{1.5}
\small
\centering
\caption{The measured differential cross-sections and the predictions from GM-VFNS and FONLL calculations for the \D meson in bins of \pt for $|\eta| < 2.5$. The statistical, systematic (excluding branching ratio) and branching ratio uncertainties are shown separately for data, while the total theoretical uncertainties are shown for GM-VFNS and FONLL.}
\begin{tabular}{c @{\hspace{0.5cm}}@{}c@{}@{(}d{1.2}@{${}\pm{}$}d{1.2}@{${}\pm{}$}d{1.2}@{${}\pm{}$}d{1.2}@{)}l r r}
\toprule
& \multicolumn{8}{c}{Differential cross-section [pb/GeV]}  \\
Range [GeV]& \multicolumn{6}{c}{{Data}} & \multicolumn{1}{c}{GM-VFNS} & \multicolumn{1}{c}{FONLL} \\
& \multicolumn{6}{c}{{$\frac{\mathrm{d}\sigma}{\mathrm{d}\pt} \pm \delta_\text{stat} \pm \delta_\text{syst} \pm \delta_\text{BR}$}} & \multicolumn{1}{c}{$\frac{\mathrm{d}\sigma}{\mathrm{d}\pt} \pm \delta_\text{theory}$} & \multicolumn{1}{c}{$\frac{\mathrm{d}\sigma}{\mathrm{d}\pt} \pm \delta_\text{theory}$} \\

\midrule

$12 < \pt < 15$   &  & 1.80  & 0.23  & 0.22 & 0.13 & $\, \times  \, 10^6$ & $2.45\numpmerr{+0.58}{-0.44} \times 10^6$ & $1.82\numpmerr{+0.54}{-0.38} \times 10^6$ \\
$15 < \pt < 20$   &  & 7.0   & 0.3   & 0.9  & 0.5  & $\, \times  \, 10^5$ & $8.5\numpmerr{+1.7}{-1.4} \times 10^5$ & $6.1\numpmerr{+1.5}{-1.1} \times 10^5$ \\
$20 < \pt < 25$   &  & 2.10  & 0.08  & 0.25 & 0.15 & $\, \times  \, 10^5$ & $2.76\numpmerr{+0.44}{-0.39} \times 10^5$ & $1.90\numpmerr{+0.39}{-0.31} \times 10^5$ \\
$25 < \pt < 30$   &  & 9.2   & 0.4   & 0.8  & 0.7  & $\, \times  \, 10^4$ & $1.10\numpmerr{+0.15}{-0.14} \times 10^5$ & $7.3\numpmerr{+1.3}{-1.1} \times 10^4$ \\
$30 < \pt < 40$   &  & 2.86  & 0.12  & 0.34 & 0.20 & $\, \times  \, 10^4$ & $3.75\numpmerr{+0.43}{-0.44} \times 10^4$ & $2.44\numpmerr{+0.38}{-0.33} \times 10^4$ \\
$40 < \pt < 50$   &  & 8.7   & 0.5   & 1.5  & 0.6  & $\, \times  \, 10^3$ & $1.08\numpmerr{+0.10}{-0.11} \times 10^4$ & $6.8\numpmerr{+0.9}{-0.8} \times 10^3$ \\
$50 < \pt < 60$   &  & 2.45  & 0.31  & 0.43 & 0.17 & $\, \times  \, 10^3$ & $3.89\numpmerr{+0.30}{-0.37} \times 10^3$ & $2.41\numpmerr{+0.29}{-0.28} \times 10^3$ \\
$60 < \pt < 80$   &  & 8.9   & 1.2   & 1.5  & 0.6  & $\, \times  \, 10^2$ & $1.21\numpmerr{+0.08}{-0.10} \times 10^3$ & $7.3\numpmerr{+0.8}{-0.8} \times 10^2$ \\
$80 < \pt < 100$  &  & 1.78  & 0.64  & 0.34 & 0.13 & $\, \times  \, 10^2$ & $3.08\numpmerr{+0.16}{-0.24} \times 10^2$ & $1.81\numpmerr{+0.19}{-0.18} \times 10^2$ \\
\bottomrule
\end{tabular}
\label{tab:XScontentdpmptoriginal}
\end{table}

\end{DIFnomarkup}


\begin{DIFnomarkup}
\begin{table}[thbp]
\renewcommand{\arraystretch}{1.5}
\small
\centering
\caption{The measured differential cross-sections and the predictions from the GM-VFNS calculation for the \Ds meson in bins of $|\eta|$ for $\SI{12}{} < \pt < \SI{100}{\GeV}$. The statistical, systematic (excluding branching ratio) and branching ratio uncertainties are shown separately for data, while the total theoretical uncertainties are shown for GM-VFNS.}
\begin{tabular}{crr}
\toprule
& \multicolumn{2}{c}{Differential cross-section [\mub]}  \\

Range & \multicolumn{1}{c}{Data} & \multicolumn{1}{c}{GM-VFNS}  \\
& \multicolumn{1}{c}{$\frac{\mathrm{d}\sigma}{\mathrm{d}|\eta|} \pm \delta_\text{stat} \pm \delta_\text{syst} \pm \delta_\text{BR}$} & \multicolumn{1}{c}{$\frac{\mathrm{d}\sigma}{\mathrm{d}|\eta|} \pm \delta_\text{theory}$ }\\

\midrule
$0.0 < |\eta| < 0.5$ & ${ 2.51 } \pm{ 0.03 } \pm{ 0.24 } \pm{0.18}$ & ${2.69}\numpmerr{+0.56}{-0.45}$ \\
$0.5 < |\eta| < 1.0$ & ${ 2.61 } \pm{ 0.03 } \pm{ 0.23 } \pm{0.19}$ & ${2.62}\numpmerr{+0.55}{-0.44}$ \\
$1.0 < |\eta| < 1.5$ & ${ 2.23 } \pm{ 0.04 } \pm{ 0.18 } \pm{0.16}$ & ${2.46}\numpmerr{+0.51}{-0.41}$ \\
$1.5 < |\eta| < 2.0$ & ${ 1.79 } \pm{ 0.03 } \pm{ 0.13 } \pm{0.13}$ & ${2.22}\numpmerr{+0.46}{-0.37}$ \\
$2.0 < |\eta| < 2.5$ & ${ 1.44 } \pm{ 0.04 } \pm{ 0.13 } \pm{0.10}$ & ${1.90}\numpmerr{+0.39}{-0.32}$ \\

\bottomrule
\end{tabular}
\label{tab:XScontentdsetaoriginal}
\end{table}

\end{DIFnomarkup}


\begin{DIFnomarkup}
\setlength{\extrarowheight}{2pt} %
\begin{table}[thbp]
\renewcommand{\arraystretch}{1.5}
\small
\centering
\caption{The measured differential cross-sections and the predictions from the GM-VFNS calculation for the \Ds meson in bins of \pt for $|\eta| < 2.5$. The statistical, systematic (excluding branching ratio) and branching ratio uncertainties are shown separately for data, while the total theoretical uncertainties are shown for GM-VFNS.}
\begin{tabular}{r @{\hspace{0.5cm}}@{}c@{}@{(}d{1.2}@{${}\pm{}$}d{1.2}@{${}\pm{}$}d{1.2}@{${}\pm{}$}d{1.2}@{)} r r}
\toprule
& \multicolumn{7}{c}{Differential cross-section [pb/GeV]}  \\
Range [GeV]& \multicolumn{6}{c}{{Data}} & \multicolumn{1}{c}{GM-VFNS} \\
& \multicolumn{6}{c}{{$\frac{\mathrm{d}\sigma}{\mathrm{d}\pt} \pm \delta_\text{stat} \pm \delta_\text{syst} \pm \delta_\text{BR}$}} & \multicolumn{1}{c}{$\frac{\mathrm{d}\sigma}{\mathrm{d}\pt} \pm \delta_\text{theory}$} \\
\midrule

$12 < \pt < 15$   &  & 8.5   & 0.4   & 0.8  & 0.6 & $\, \times  \, 10^5$ & $1.02\numpmerr{+0.24}{-0.18} \times 10^6$ \\
$15 < \pt < 20$   &  & 3.04  & 0.06  & 0.29 & 0.22 & $\, \times  \, 10^5$ & $3.62\numpmerr{+0.71}{-0.59} \times 10^5$ \\
$20 < \pt < 25$   &  & 1.01  & 0.01  & 0.08 & 0.07 & $\, \times  \, 10^5$ & $1.19\numpmerr{+0.19}{-0.17} \times 10^5$ \\
$25 < \pt < 30$   &  & 4.20  & 0.06  & 0.31 & 0.31 & $\, \times  \, 10^4$ & $4.75\numpmerr{+0.65}{-0.62} \times 10^4$ \\
$30 < \pt < 40$   &  & 1.45  & 0.02  & 0.12 & 0.11 & $\, \times  \, 10^4$ & $1.64\numpmerr{+0.19}{-0.19} \times 10^4$ \\
$40 < \pt < 50$   &  & 3.82  & 0.09  & 0.36 & 0.28 & $\, \times  \, 10^3$ & $4.74\numpmerr{+0.45}{-0.50} \times 10^3$ \\
$50 < \pt < 60$   &  & 1.38  & 0.05  & 0.13 & 0.10 & $\, \times  \, 10^3$ & $1.72\numpmerr{+0.14}{-0.17} \times 10^3$ \\
$60 < \pt < 80$   &  & 4.55  & 0.21  & 0.44 & 0.33 & $\, \times  \, 10^2$ & $5.39\numpmerr{+0.37}{-0.47} \times 10^2$ \\
$80 < \pt < 100$  &  & 8.1   & 0.9   & 0.9  & 0.6 & $\, \times  \, 10^1$ & $1.38\numpmerr{+0.08}{-0.11} \times 10^2$ \\
\bottomrule
\end{tabular}
\label{tab:XScontentdsptoriginal}
\end{table}

\end{DIFnomarkup}


\FloatBarrier
\begin{figure}[thbp]
\centering
\subfloat[][]{\includegraphics[width=0.5\textwidth]{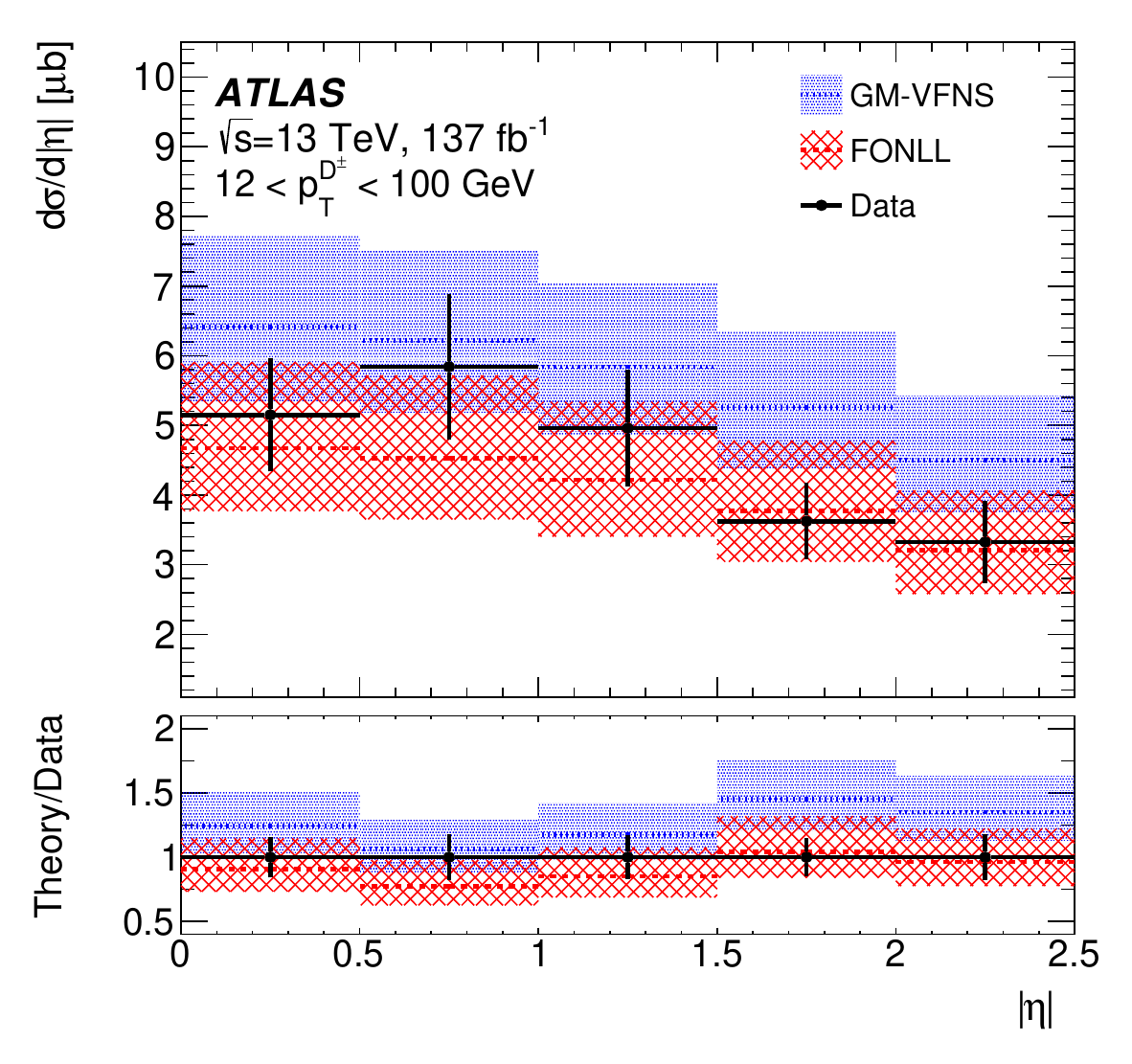}}
\subfloat[][]{\includegraphics[width=0.5\textwidth]{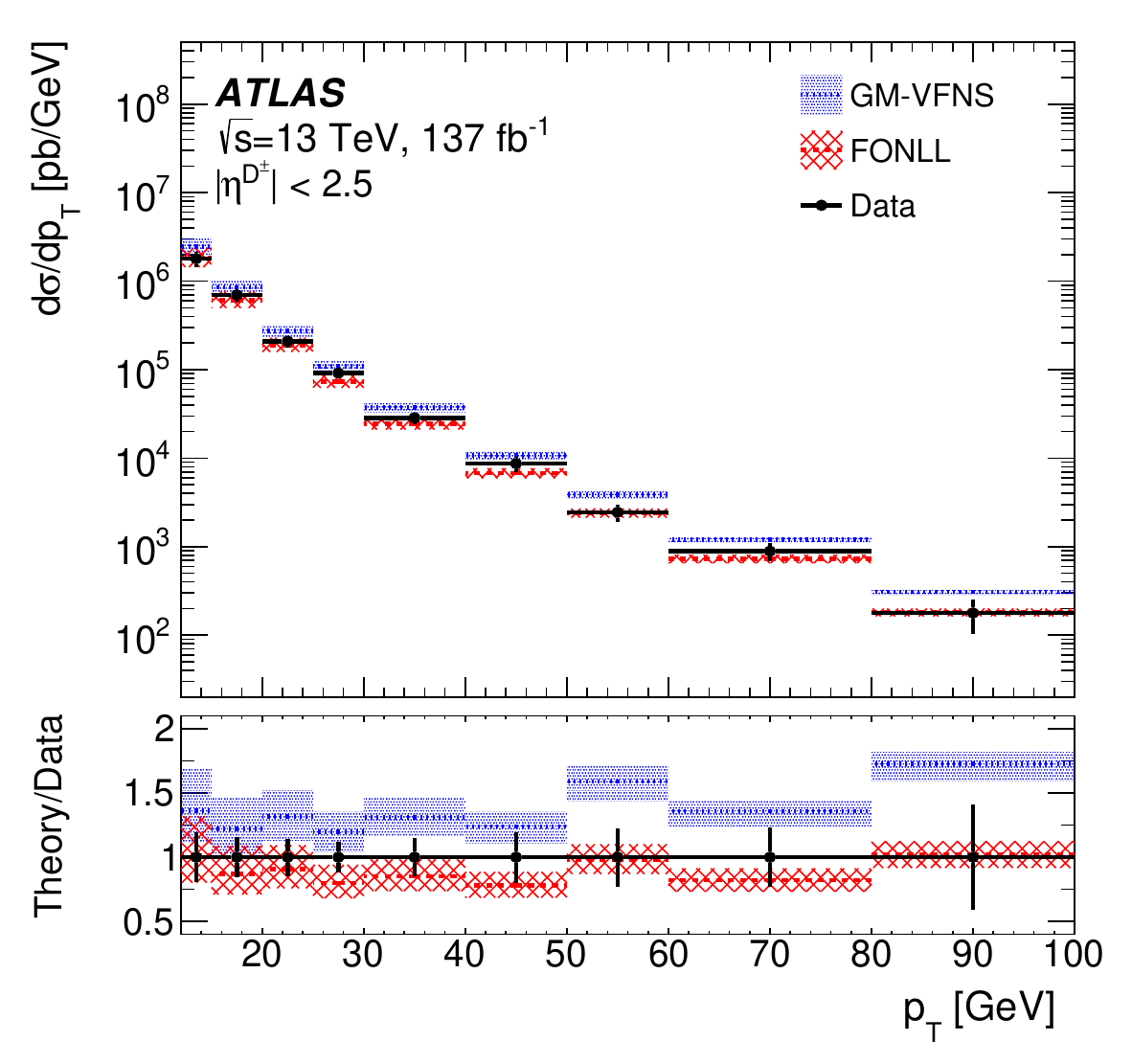}}\\
\subfloat[][]{\includegraphics[width=0.5\textwidth]{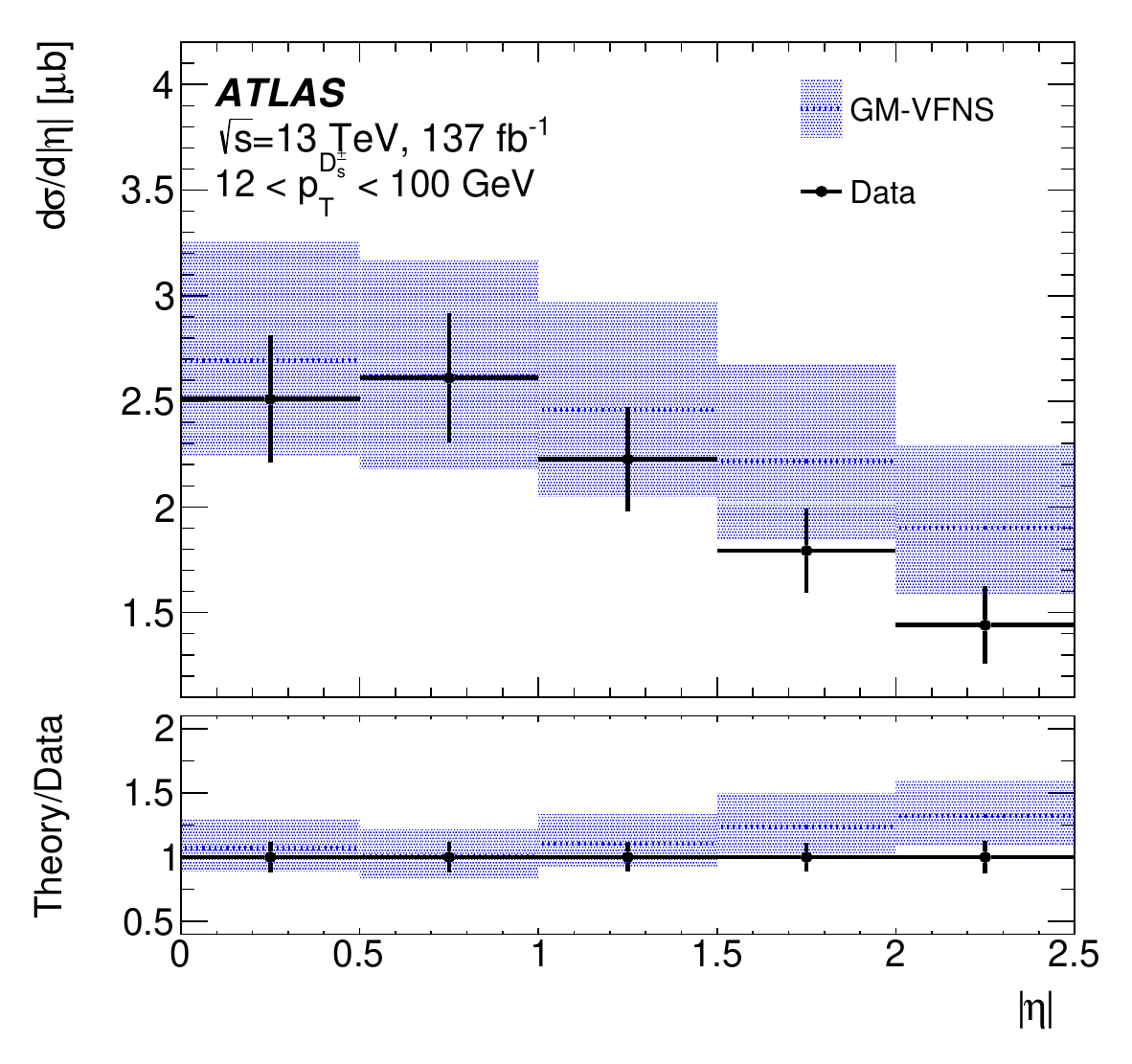}}
\subfloat[][]{\includegraphics[width=0.5\textwidth]{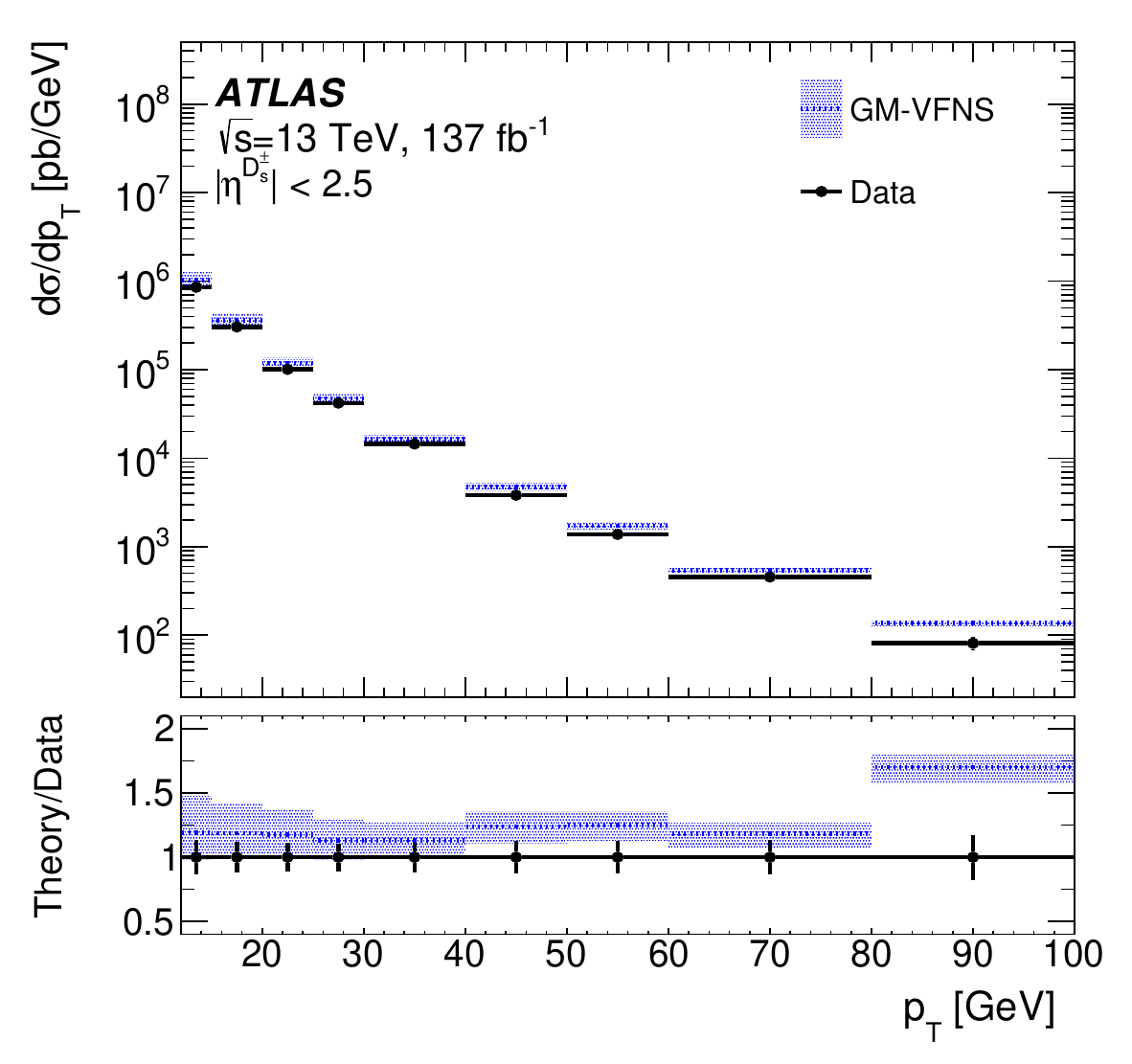}}
\caption{Differential cross-sections for \D mesons in bins of (a) $|\eta|$ and (b) \pt, and for \Ds mesons in bins of (c) $|\eta|$ and (d) \pt. The measured values with the statistical and systematic uncertainty are represented by the data points, while the theory predictions from GM-VFNS and FONLL are shown as dotted and dashed lines, respectively; the uncertainties on the theory predictions are represented by the hatched areas. The theory predictions are also divided by the measured values to obtain the ratios between the predictions and the measurements shown in the bottom panels.}
\label{fig:dXS_dq}
\end{figure}
\FloatBarrier

\begin{DIFnomarkup}
\begin{table}[!t]
\renewcommand{\arraystretch}{1.5}
\footnotesize
\centering
\caption{Inclusive \D and \Ds meson production cross-sections in different fiducial volumes defined by $|\eta|<2.5$ and different \pT regions. For the ATLAS measurements, the statistical, systematic (excluding branching ratio) and branching ratio uncertainties are shown separately; for the theoretical predictions, the total theoretical uncertainties are shown.}
\begin{tabular}{cccccc}
\toprule
& \multicolumn{3}{c}{\D inclusive fiducial cross-section at \sqrtsthirteen[nb]} \\
Fiducial volume & ATLAS                & GM-VFNS                     &FONLL        &\\
&$\sigma\pm\delta_\text{stat}\pm\delta_\text{syst}\pm\delta_\text{BR}$& $\sigma\pm\delta_\text{theory}$ &$\sigma\pm\delta_\text{theory}$ & \\
\midrule
$12<\pt<100~\GeV, |\eta|<2.5$ &	$10\,800 \pm 900 \pm 1\,300 \pm 800$         &  $14\,100^{+2\,900}_{-2\,300}$  &   $10\,200^{+2\,300}_{-1\,700}$ &\\%
$15<\pt<100~\GeV, |\eta|<2.5$ &	$5\,430      \pm 550 \pm 680 \pm 390$      &  $6\,800^{+1\,200}_{-1\,000}$   &    $4\,730^{+900}_{-700}$&  \\%
$20<\pt<100~\GeV, |\eta|<2.5$ &	$1\,930      \pm 160 \pm 220 \pm 140$      &  $2\,480^{+350}_{-330}$   &    $1\,670^{+260}_{-220}$&  \\%

\toprule
& \multicolumn{4}{c}{\Ds inclusive fiducial cross-section at \sqrtsthirteen[nb]}  \\
Fiducial volume & ATLAS     & GM-VFNS        \\
&$\sigma\pm\delta_\text{stat}\pm\delta_\text{syst}\pm\delta_\text{BR}$& $\sigma\pm\delta_\text{theory}$ &\\
\midrule
$12<\pt<100~\GeV, |\eta|<2.5$  &	$5\,000 \pm 360 \pm 470 \pm 360$       &  $5\,900^{+1\,200}_{-1\,000}$ &  \\
$15<\pt<100~\GeV, |\eta|<2.5$  &	$2\,440 \pm 190 \pm 220 \pm 180$       &  $2\,880^{+510}_{-440}$  &  \\
$20<\pt<100~\GeV, |\eta|<2.5$  &    $920 \pm 60 \pm 80 \pm 70$       &  $1\,070^{+150}_{-140}$        &\\

\bottomrule
\end{tabular}
\label{tab:xsec:Ds}
\end{table}

\end{DIFnomarkup}


\end{DIFnomarkup}

For the \D meson, good agreement is observed at low \pt for both the GM-VFNS and FONLL predictions; the FONLL prediction is in general slightly lower than the measured differential cross-sections, while the GM-VFNS prediction is slightly higher.
For higher \pt, the GM-VFNS prediction gives a larger value than the measured differential cross-section, while the FONLL prediction is still consistent with the measured values.
When comparing in bins of $|\eta|$, good agreement is observed for both predictions, which is consistent with the observation in the low \pt bins.

For the \Ds meson, only the GM-VFNS prediction is available for comparison. Similar behaviour is observed; the GM-VFNS prediction gives a larger value than the measured differential cross-section for both $|\eta|$ and \pt, with a trend towards a larger deviation at higher \pt.

The results are compared with those measured with the \pp collision data collected at a centre-of-mass energy of \sqrtseven in 2010~\cite{BPHY-2013-04}. Those measurements were performed using the $\D \rightarrow K^\mp  \pi^\pm \pi^\pm$ and $\Ds \rightarrow \phi(K^+K^-)\pi^\pm$ decays for $|\eta|<2.1$.
To perform a consistent comparison, the difference between the fiducial volumes, particularly in $|\eta|$, must be taken into account. Since the differential cross-section in bins of $|\eta|$ is rather flat and well modelled by MC simulation, the ratio of the fiducial volume of $|\eta|<2.1$ to $|\eta|<2.5$ is derived from MC simulation with a negligible systematic uncertainty.
Table~\ref{tab:xsec:DsvsR1} shows the comparison between the fiducial cross-sections measured at \sqrtsthirteen and \sqrtseven. For the \Ds meson, the measurement has a 12\% uncertainty, which is an improvement over the 20\% uncertainty in the previous measurement.
For the \D meson, the measurement has a 14\% uncertainty, which is larger than the previous uncertainty of 11\%, mainly due to the background uncertainty driven by the low signal yield.

For both the ATLAS measurements, and the theory predictions, ratios between \sqrtseven and \sqrtsthirteen are also computed. The statistical uncertainties in the two ATLAS measurements are independent because of the different data samples used; the systematic uncertainties are also assumed to be uncorrelated, because the two measurements used different decay channels with all main systematic uncertainty contributions being different. For FONLL and GM-VFNS, the uncertainties are assumed to be fully correlated between \sqrtseven and \sqrtsthirteen, as they are evaluated mainly from the variations of QCD scale uncertainties in a coherent way. In Table~\ref{tab:xsec:DsvsR1}, it can be seen that the ratios obtained by GM-VFNS and FONLL are both consistent with the data within the uncertainties. There is, however, a significant difference between the ratios predicted by GM-VFNS and FONLL.

\begin{DIFnomarkup}
\begin{table}[!t]
\renewcommand{\arraystretch}{1.5}
\footnotesize
\centering
\caption{Inclusive \D and \Ds meson production cross-sections in a fiducial volume defined by $|\eta|<2.1$ and $20<\pt<100~\GeV$ at \sqrtseven and \sqrtsthirteen. The total uncertainties are shown for both the ATLAS measurements and the theoretical predictions. The ratios of \sqrtsthirteen values over \sqrtseven values are also computed by assuming the uncertainties are independent for different centre-of-mass energies.} %
\begin{tabular}{cccc}
\toprule

& \multicolumn{3}{c}{\D inclusive fiducial cross-section [nb]} \\

& ATLAS                & GM-VFNS                     &FONLL        \\
&$\sigma\pm\delta_\text{total}$& $\sigma\pm\delta_\text{theory}$ &$\sigma\pm\delta_\text{theory}$  \\

\midrule

\sqrtsthirteen &	$1\,690      \pm 270$  & $2\,200^{+310}_{-290}$   & $1\,480^{+230}_{-190}$   \\%
\sqrtseven &	   $888 \pm 97$   &  $980^{+120}_{-150}$  &   $620^{+100}_{-80}$\\%
\midrule
Ratio (\SI{13}{\TeV}/\SI{7}{\TeV}) &	  $1.9 \pm 0.4$    &  $2.24 \pm 0.04$ & $2.38 \pm 0.01$     \\%
\toprule
& \multicolumn{3}{c}{\Ds inclusive fiducial cross-section [nb]}  \\
& ATLAS                & GM-VFNS                     &       \\
&$\sigma\pm\delta_\text{total}$& $\sigma\pm\delta_\text{theory}$  \\

\midrule

\sqrtsthirteen &	$810 \pm 100$  &  $950^{+140}_{-130}$  &    \\%
\sqrtseven &	   $510\pm 100$   &  $470^{+56}_{-69}$  &   \\%
\midrule
Ratio (\SI{13}{\TeV}/\SI{7}{\TeV}) &	  $1.6 \pm 0.4$    &  $2.02 \pm 0.05$ &      \\%

\bottomrule
\end{tabular}
\label{tab:xsec:DsvsR1}
\end{table}
\end{DIFnomarkup}



%

%
%

%
%
%
\FloatBarrier
%

\section{Conclusions}

\FloatBarrier
The production of \D and \Ds charmed mesons is measured in the kinematic region $\SI{12}{} < \pT < \SI{100}{\GeV}$ and $|\eta| < 2.5$ with the ATLAS detector at the LHC using \lumi of \sqsthirt \pp collision data.
The differential cross-sections in bins of \pt and $|\eta|$ for \D and \Ds meson production are determined and compared with the available NLO QCD predictions.
The fiducial cross-sections for \D and \Ds meson production are also presented; the values are compared with the predictions and with the measurement at \sqrtseven.
For the \Ds meson, this is the first measurement of the differential cross-section reported by the ATLAS Collaboration, and the first time such a measurement has been reported up to transverse momenta of \SI{100}{\GeV}, providing a benchmark for theoretical calculations in a previously unexplored kinematic space.
The obtained cross-sections are compared with the predictions from the GM-VFNS and FONLL calculations, and the results are mostly consistent within the uncertainties, with slight deviation towards high-\pt regions.


\section*{Acknowledgements}

%
%

%
%

We thank CERN for the very successful operation of the LHC and its injectors, as well as the support staff at
CERN and at our institutions worldwide without whom ATLAS could not be operated efficiently.

The crucial computing support from all WLCG partners is acknowledged gratefully, in particular from CERN, the ATLAS Tier-1 facilities at TRIUMF/SFU (Canada), NDGF (Denmark, Norway, Sweden), CC-IN2P3 (France), KIT/GridKA (Germany), INFN-CNAF (Italy), NL-T1 (Netherlands), PIC (Spain), RAL (UK) and BNL (USA), the Tier-2 facilities worldwide and large non-WLCG resource providers. Major contributors of computing resources are listed in Ref.~\cite{ATL-SOFT-PUB-2025-001}.

We gratefully acknowledge the support of ANPCyT, Argentina; YerPhI, Armenia; ARC, Australia; BMWFW and FWF, Austria; ANAS, Azerbaijan; CNPq and FAPESP, Brazil; NSERC, NRC and CFI, Canada; CERN; ANID, Chile; CAS, MOST and NSFC, China; Minciencias, Colombia; MEYS CR, Czech Republic; DNRF and DNSRC, Denmark; IN2P3-CNRS and CEA-DRF/IRFU, France; SRNSFG, Georgia; BMBF, HGF and MPG, Germany; GSRI, Greece; RGC and Hong Kong SAR, China; ISF and Benoziyo Center, Israel; INFN, Italy; MEXT and JSPS, Japan; CNRST, Morocco; NWO, Netherlands; RCN, Norway; MNiSW, Poland; FCT, Portugal; MNE/IFA, Romania; MSTDI, Serbia; MSSR, Slovakia; ARIS and MVZI, Slovenia; DSI/NRF, South Africa; MICIU/AEI, Spain; SRC and Wallenberg Foundation, Sweden; SERI, SNSF and Cantons of Bern and Geneva, Switzerland; NSTC, Taipei; TENMAK, T\"urkiye; STFC/UKRI, United Kingdom; DOE and NSF, United States of America.

Individual groups and members have received support from BCKDF, CANARIE, CRC and DRAC, Canada; CERN-CZ, FORTE and PRIMUS, Czech Republic; COST, ERC, ERDF, Horizon 2020, ICSC-NextGenerationEU and Marie Sk{\l}odowska-Curie Actions, European Union; Investissements d'Avenir Labex, Investissements d'Avenir Idex and ANR, France; DFG and AvH Foundation, Germany; Herakleitos, Thales and Aristeia programmes co-financed by EU-ESF and the Greek NSRF, Greece; BSF-NSF and MINERVA, Israel; NCN and NAWA, Poland; La Caixa Banking Foundation, CERCA Programme Generalitat de Catalunya and PROMETEO and GenT Programmes Generalitat Valenciana, Spain; G\"{o}ran Gustafssons Stiftelse, Sweden; The Royal Society and Leverhulme Trust, United Kingdom.

In addition, individual members wish to acknowledge support from Armenia: Yerevan Physics Institute (FAPERJ); CERN: European Organization for Nuclear Research (CERN PJAS); Chile: Agencia Nacional de Investigaci\'on y Desarrollo (FONDECYT 1230812, FONDECYT 1230987, FONDECYT 1240864); China: Chinese Ministry of Science and Technology (MOST-2023YFA1605700, MOST-2023YFA1609300), National Natural Science Foundation of China (NSFC - 12175119, NSFC 12275265, NSFC-12075060); Czech Republic: Czech Science Foundation (GACR - 24-11373S), Ministry of Education Youth and Sports (ERC-CZ-LL2327, FORTE CZ.02.01.01/00/22\_008/0004632), PRIMUS Research Programme (PRIMUS/21/SCI/017); EU: H2020 European Research Council (ERC - 101002463); European Union: European Research Council (ERC - 948254, ERC 101089007), Horizon 2020 Framework Programme (MUCCA - CHIST-ERA-19-XAI-00), European Union, Future Artificial Intelligence Research (FAIR-NextGenerationEU PE00000013), Italian Center for High Performance Computing, Big Data and Quantum Computing (ICSC, NextGenerationEU); France: Agence Nationale de la Recherche (ANR-20-CE31-0013, ANR-21-CE31-0013, ANR-21-CE31-0022, ANR-22-EDIR-0002), Investissements d'Avenir Labex (ANR-11-LABX-0012); Germany: Baden-Württemberg Stiftung (BW Stiftung-Postdoc Eliteprogramme), Deutsche Forschungsgemeinschaft (DFG - 469666862, DFG - CR 312/5-2); Italy: Istituto Nazionale di Fisica Nucleare (ICSC, NextGenerationEU), Ministero dell'Università e della Ricerca (NextGenEU PRIN20223N7F8K M4C2.1.1); Japan: Japan Society for the Promotion of Science (JSPS KAKENHI JP22H01227, JSPS KAKENHI JP22H04944, JSPS KAKENHI JP22KK0227, JSPS KAKENHI JP23KK0245); Netherlands: Netherlands Organisation for Scientific Research (NWO Veni 2020 - VI.Veni.202.179); Norway: Research Council of Norway (RCN-314472); Poland: Ministry of Science and Higher Education (IDUB AGH, POB8, D4 no 9722), Polish National Agency for Academic Exchange (PPN/PPO/2020/1/00002/U/00001), Polish National Science Centre (NCN 2021/42/E/ST2/00350, NCN OPUS nr 2022/47/B/ST2/03059, NCN UMO-2019/34/E/ST2/00393, NCN \& H2020 MSCA 945339, UMO-2020/37/B/ST2/01043, UMO-2021/40/C/ST2/00187, UMO-2022/47/O/ST2/00148, UMO-2023/49/B/ST2/04085, UMO-2023/51/B/ST2/00920); Portugal: Foundation for Science and Technology (FCT); Slovenia: Slovenian Research Agency (ARIS grant J1-3010); Spain: Generalitat Valenciana (Artemisa, FEDER, IDIFEDER/2018/048), Ministry of Science and Innovation (MCIN \& NextGenEU PCI2022-135018-2, MICIN \& FEDER PID2021-125273NB, RYC2019-028510-I, RYC2020-030254-I, RYC2021-031273-I, RYC2022-038164-I), PROMETEO and GenT Programmes Generalitat Valenciana (CIDEGENT/2019/027); Sweden: Carl Trygger Foundation (Carl Trygger Foundation CTS 22:2312), Swedish Research Council (Swedish Research Council 2023-04654, VR 2018-00482, VR 2021-03651, VR 2022-03845, VR 2022-04683, VR 2023-03403), Knut and Alice Wallenberg Foundation (KAW 2018.0157, KAW 2018.0458, KAW 2019.0447, KAW 2022.0358); Switzerland: Swiss National Science Foundation (SNSF - PCEFP2\_194658); United Kingdom: Leverhulme Trust (Leverhulme Trust RPG-2020-004), Royal Society (NIF-R1-231091); United States of America: U.S. Department of Energy (ECA DE-AC02-76SF00515), Neubauer Family Foundation.

%
%


\printbibliography

\clearpage
\input{atlas_authlist}

\clearpage

\end{document}

%% file: atlas_authlist.tex
 
\begin{flushleft}
\hypersetup{urlcolor=black}
{\Large The ATLAS Collaboration}

\bigskip

\AtlasOrcid[0000-0002-6665-4934]{G.~Aad}$^\textrm{\scriptsize 104}$,
\AtlasOrcid[0000-0001-7616-1554]{E.~Aakvaag}$^\textrm{\scriptsize 17}$,
\AtlasOrcid[0000-0002-5888-2734]{B.~Abbott}$^\textrm{\scriptsize 123}$,
\AtlasOrcid[0000-0002-0287-5869]{S.~Abdelhameed}$^\textrm{\scriptsize 119a}$,
\AtlasOrcid[0000-0002-1002-1652]{K.~Abeling}$^\textrm{\scriptsize 56}$,
\AtlasOrcid[0000-0001-5763-2760]{N.J.~Abicht}$^\textrm{\scriptsize 50}$,
\AtlasOrcid[0000-0002-8496-9294]{S.H.~Abidi}$^\textrm{\scriptsize 30}$,
\AtlasOrcid[0009-0003-6578-220X]{M.~Aboelela}$^\textrm{\scriptsize 45}$,
\AtlasOrcid[0000-0002-9987-2292]{A.~Aboulhorma}$^\textrm{\scriptsize 36e}$,
\AtlasOrcid[0000-0001-5329-6640]{H.~Abramowicz}$^\textrm{\scriptsize 155}$
% ...
\bigskip
\\

% ...
$^{17}$Department for Physics and Technology, University of Bergen, Bergen; Norway.\\
% ...
$^{30}$Physics Department, Brookhaven National Laboratory, Upton NY; United States of America.\\
% ...
$^{36}$$^{(a)}$Facult\'e des Sciences Ain Chock, Universit\'e Hassan II de Casablanca;$^{(b)}$Facult\'{e} des Sciences, Universit\'{e} Ibn-Tofail, K\'{e}nitra;$^{(c)}$Facult\'e des Sciences Semlalia, Universit\'e Cadi Ayyad, LPHEA-Marrakech;$^{(d)}$LPMR, Facult\'e des Sciences, Universit\'e Mohamed Premier, Oujda;$^{(e)}$Facult\'e des sciences, Universit\'e Mohammed V, Rabat;$^{(f)}$Institute of Applied Physics, Mohammed VI Polytechnic University, Ben Guerir; Morocco.\\
% ...
$^{45}$Physics Department, Southern Methodist University, Dallas TX; United States of America.\\
% ...
$^{50}$Fakult\"{a}t Physik , Technische Universit{\"a}t Dortmund, Dortmund; Germany.\\
% ...
$^{56}$II. Physikalisches Institut, Georg-August-Universit\"{a}t G\"ottingen, G\"ottingen; Germany.\\
% ...
$^{104}$CPPM, Aix-Marseille Universit\'e, CNRS/IN2P3, Marseille; France.\\
% ...
$^{119}$$^{(a)}$New York University Abu Dhabi, Abu Dhabi;$^{(b)}$United Arab Emirates University, Al Ain; United Arab Emirates.\\
% ...
$^{123}$Homer L. Dodge Department of Physics and Astronomy, University of Oklahoma, Norman OK; United States of America.\\
% ...
$^{155}$Raymond and Beverly Sackler School of Physics and Astronomy, Tel Aviv University, Tel Aviv; Israel.\\
% ...
$^{a}$ Also Affiliated with an institute covered by a cooperation agreement with CERN.\\
$^{b}$ Also at An-Najah National University, Nablus; Palestine.\\
$^{c}$ Also at Borough of Manhattan Community College, City University of New York, New York NY; United States of America.\\
$^{d}$ Also at Center for High Energy Physics, Peking University; China.\\
$^{e}$ Also at Center for Interdisciplinary Research and Innovation (CIRI-AUTH), Thessaloniki; Greece.\\
$^{f}$ Also at CERN, Geneva; Switzerland.\\
% ...

\end{flushleft}


%% file: ANA-BPHY-2018-10-INT1.bib
@Book{Voigt1912,

    author = "Voigt, Woldemar",

    title = "On the law of intensity distribution within the lines of a gas spectrum",

    year = "1912",

    address = "Munich",

    series = "Meeting reports",

    volume = "1912,25",

    url = "https://publikationen.badw.de/de/003395768",

}

@article{Bukin:2007zha,
    author = "Bukin, A. D.",
    title = "{Fitting function for asymmetric peaks}",
    eprint = "0711.4449",
    archivePrefix = "arXiv",
    primaryClass = "physics.data-an",
    month = "11",
    year = "2007"
}

@article{ParticleDataGroup:2024cfk,

    author = "Navas, S. and others",

    collaboration = "Particle Data Group",

    title = "{Review of particle physics}",

    doi = "10.1103/PhysRevD.110.030001",

    journal = "Phys. Rev. D",

    volume = "110",

    number = "3",

    pages = "030001",

    year = "2024"

}

@article{LHCb:2015swx,
    author = "{LHCb Collaboration}",
    title = "{Measurements of prompt charm production cross-sections in $pp$ collisions at $ \sqrt{s}=13 $ TeV}",
    eprint = "1510.01707",
    archivePrefix = "arXiv",
    primaryClass = "hep-ex",
    reportNumber = "LHCB-PAPER-2015-041, CERN-PH-EP-2015-272",
    doi = "10.1007/JHEP03(2016)159",
    journal = "JHEP",
    volume = "03",
    pages = "159",
    year = "2016",
}

@article{ALICE:2023sgl,
    author = "{ALICE Collaboration}",
    collaboration = "ALICE",
    title = "{Charm production and fragmentation fractions at midrapidity in pp collisions at $ \sqrt{\textrm{s}} $ = 13 TeV}",
    eprint = "2308.04877",
    archivePrefix = "arXiv",
    primaryClass = "hep-ex",
    reportNumber = "CERN-EP-2023-162",
    doi = "10.1007/JHEP12(2023)086",
    journal = "JHEP",
    volume = "12",
    pages = "086",
    year = "2023"
}

@article{ALICE:2024xln,
    author = "{ALICE Collaboration}",
    collaboration = "ALICE",
    title = "{Measurement of beauty-quark production in pp collisions at $ \sqrt{s} $ = 13 TeV via non-prompt D mesons}",
    eprint = "2402.16417",
    archivePrefix = "arXiv",
    primaryClass = "hep-ex",
    reportNumber = "CERN-EP-2024-055",
    doi = "10.1007/JHEP10(2024)110",
    journal = "JHEP",
    volume = "10",
    pages = "110",
    year = "2024"
}

@article{ALICE:2012inj,
    author = "{ALICE Collaboration}",
    collaboration = "ALICE",
    title = "{Measurement of charm production at central rapidity in proton-proton collisions at $\sqrt{s}=2.76$ TeV}",
    eprint = "1205.4007",
    archivePrefix = "arXiv",
    primaryClass = "hep-ex",
    reportNumber = "CERN-PH-EP-2012-133",
    doi = "10.1007/JHEP07(2012)191",
    journal = "JHEP",
    volume = "07",
    pages = "191",
    year = "2012"
}

@article{ALICE:2017olh,
    author = "{ALICE Collaboration}",
    collaboration = "ALICE",
    title = "{Measurement of D-meson production at mid-rapidity in pp collisions at ${\sqrt{s}=7}$  TeV}",
    eprint = "1702.00766",
    archivePrefix = "arXiv",
    primaryClass = "hep-ex",
    reportNumber = "CERN-EP-2017-020",
    doi = "10.1140/epjc/s10052-017-5090-4",
    journal = "Eur. Phys. J. C",
    volume = "77",
    number = "8",
    pages = "550",
    year = "2017"
}

@article{ALICE:2019nxm,
    author = "{ALICE Collaboration}",
    collaboration = "ALICE",
    title = "{Measurement of ${{\mathrm{D}}^0}$ , ${{\mathrm{D}}^+}$ , ${{\mathrm{D}}^{*+}}$ and ${{\mathrm{D}}^+_{\mathrm{s}}}$ production in pp collisions at ${\sqrt{{\textit{s}}}~=~5.02~{\text {TeV}}}$ with ALICE}",
    eprint = "1901.07979",
    archivePrefix = "arXiv",
    primaryClass = "nucl-ex",
    reportNumber = "CERN-EP-2019-004",
    doi = "10.1140/epjc/s10052-019-6873-6",
    journal = "Eur. Phys. J. C",
    volume = "79",
    number = "5",
    pages = "388",
    year = "2019"
}

@article{ALICE:2021mgk,
    author = "{ALICE Collaboration}",
    collaboration = "ALICE",
    title = "{Measurement of beauty and charm production in pp collisions at $ \sqrt{s} $ = 5.02 TeV via non-prompt and prompt D mesons}",
    eprint = "2102.13601",
    archivePrefix = "arXiv",
    primaryClass = "nucl-ex",
    reportNumber = "CERN-EP-2021-034",
    doi = "10.1007/JHEP05(2021)220",
    journal = "JHEP",
    volume = "05",
    pages = "220",
    year = "2021"
}

@article{LHCb:2016ikn,
    author = "{LHCb Collaboration}",
    collaboration = "LHCb",
    title = "{Measurements of prompt charm production cross-sections in pp collisions at $ \sqrt{s}=5 $ TeV}",
    eprint = "1610.02230",
    archivePrefix = "arXiv",
    primaryClass = "hep-ex",
    reportNumber = "CERN-EP-2016-244, LHCB-PAPER-2016-042",
    doi = "10.1007/JHEP06(2017)147",
    journal = "JHEP",
    volume = "06",
    pages = "147",
    year = "2017"
}

@article{LHCb:2013xam,
    author = "{LHCb Collaboration}",
    collaboration = "LHCb",
    title = "{Prompt charm production in pp collisions at sqrt(s)=7 TeV}",
    eprint = "1302.2864",
    archivePrefix = "arXiv",
    primaryClass = "hep-ex",
    reportNumber = "LHCB-PAPER-2012-041, CERN-PH-EP-2013-009",
    doi = "10.1016/j.nuclphysb.2013.02.010",
    journal = "Nucl. Phys. B",
    volume = "871",
    pages = "1--20",
    year = "2013"
}

@article{Cacciari2012,
    author = "Cacciari, Matteo and Frixione, Stefano and Houdeau, Nicolas and Mangano, Michelangelo L. and Nason, Paolo and Ridolfi, Giovanni",
    title = "{Theoretical predictions for charm and bottom production at the LHC}",
    eprint = "1205.6344",
    archivePrefix = "arXiv",
    primaryClass = "hep-ph",
    reportNumber = "CERN-PH-TH-2011-227",
    doi = "10.1007/JHEP10(2012)137",
    journal = "JHEP",
    volume = "10",
    pages = "137",
    year = "2012"
}

@article{Cacciari:1998it,
    author = "Cacciari, Matteo and Greco, Mario and Nason, Paolo",
    title = "{The $p_\text{T}$ spectrum in heavy-flavour hadroproduction}",
    eprint = "hep-ph/9803400",
    archivePrefix = "arXiv",
    reportNumber = "CERN-TH-98-77, LPTHE-ORSAY-98-11, IFUM-613-FT, LNF-98-008-P",
    doi = "10.1088/1126-6708/1998/05/007",
    journal = "JHEP",
    volume = "05",
    pages = "007",
    year = "1998"
}

@article{Cacciari:2001td,
    author = "Cacciari, Matteo and Frixione, Stefano and Nason, Paolo",
    title = "{The $p_\text{T}$ spectrum in heavy-flavour photoproduction}",
    eprint = "hep-ph/0102134",
    archivePrefix = "arXiv",
    reportNumber = "BICOCCA-FT-01-01, GEF-TH-2-01, YITP-SB-01-01",
    doi = "10.1088/1126-6708/2001/03/006",
    journal = "JHEP",
    volume = "03",
    pages = "006",
    year = "2001"
}

@article{Cacciari:2015fta,
    author = "Cacciari, Matteo and Mangano, Michelangelo L. and Nason, Paolo",
    title = "{Gluon PDF constraints from the ratio of forward heavy-quark production at the LHC at $\sqrt{S}=7$ and 13 TeV}",
    eprint = "1507.06197",
    archivePrefix = "arXiv",
    primaryClass = "hep-ph",
    reportNumber = "CERN-PH-TH-2015-171",
    doi = "10.1140/epjc/s10052-015-3814-x",
    journal = "Eur. Phys. J. C",
    volume = "75",
    number = "12",
    pages = "610",
    year = "2015"
}

@article{PhysRevD.71.094013,
    author = "Kniehl, Bernd A. and Kramer, Gustav",
    title = "{$D^0$, $D^+$, $D^+_s$, and $\Lambda^+_c$ fragmentation functions from CERN LEP1}",
    eprint = "hep-ph/0504058",
    archivePrefix = "arXiv",
    reportNumber = "DESY-05-054, INT-ACK-05-34",
    doi = "10.1103/PhysRevD.71.094013",
    journal = "Phys. Rev. D",
    volume = "71",
    pages = "094013",
    year = "2005"
}

@article{Benzke2017,
    author = "Benzke, M. and Garzelli, M. V. and Kniehl, B. and Kramer, G. and Moch, S. and Sigl, G.",
    title = "{Prompt neutrinos from atmospheric charm in the general-mass variable-flavor-number scheme}",
    eprint = "1705.10386",
    archivePrefix = "arXiv",
    primaryClass = "hep-ph",
    reportNumber = "DESY-17-073",
    doi = "10.1007/JHEP12(2017)021",
    journal = "JHEP",
    volume = "12",
    pages = "021",
    year = "2017"
}

@article{PhysRevD.71.014018,
    author = "Kniehl, B. A. and Kramer, G. and Schienbein, I. and Spiesberger, H.",
    title = "{Inclusive $D^{*\pm}$ production in \pp collisions with massive charm quarks}",
    eprint = "hep-ph/0410289",
    archivePrefix = "arXiv",
    reportNumber = "DESY-04-172, MZ-TH-04-07",
    doi = "10.1103/PhysRevD.71.014018",
    journal = "Phys. Rev. D",
    volume = "71",
    pages = "014018",
    year = "2005"
}

@article{PhysRevLett.96.012001,
    author = "Kniehl, B. A. and Kramer, G. and Schienbein, I. and Spiesberger, H.",
    title = "{Reconciling open charm production at the Fermilab Tevatron with QCD}",
    eprint = "hep-ph/0508129",
    archivePrefix = "arXiv",
    reportNumber = "DESY-05-146, MZ-TH-05-16",
    doi = "10.1103/PhysRevLett.96.012001",
    journal = "Phys. Rev. Lett.",
    volume = "96",
    pages = "012001",
    year = "2006"
}

@article{PhysRevD.79.094009,
    author = "Kniehl, Bernd A. and Kramer, Gustav and Schienbein, Ingo and Spiesberger, Hubert",
    title = "{Open charm hadroproduction and the charm content of the proton}",
    eprint = "0901.4130",
    archivePrefix = "arXiv",
    primaryClass = "hep-ph",
    reportNumber = "DESY-09-008, MZ-TH-09-03, LPSC-09-17",
    doi = "10.1103/PhysRevD.79.094009",
    journal = "Phys. Rev. D",
    volume = "79",
    pages = "094009",
    year = "2009"
}

@article{Gladilin:2014tba,
    author = "Gladilin, Leonid",
    title = "{Fragmentation fractions of $c$ and $b$ quarks into charmed hadrons at LEP}",
    eprint = "1404.3888",
    archivePrefix = "arXiv",
    primaryClass = "hep-ex",
    doi = "10.1140/epjc/s10052-014-3250-3",
    journal = "Eur. Phys. J. C",
    volume = "75",
    number = "1",
    pages = "19",
    year = "2015"
}

@article{Kneesch:2007ey,
    author = "Kneesch, T. and Kniehl, B. A. and Kramer, G. and Schienbein, I.",
    title = "{Charmed-meson fragmentation functions with finite-mass corrections}",
    eprint = "0712.0481",
    archivePrefix = "arXiv",
    primaryClass = "hep-ph",
    reportNumber = "DESY-07-215, LPSC-07-131",
    doi = "10.1016/j.nuclphysb.2008.02.015",
    journal = "Nucl. Phys. B",
    volume = "799",
    pages = "34--59",
    year = "2008"
}

@article{Kniehl:2012ti,
    author = "Kniehl, B. A. and Kramer, G. and Schienbein, I. and Spiesberger, H.",
    title = "{Inclusive charmed-meson production at the CERN LHC}",
    eprint = "1202.0439",
    archivePrefix = "arXiv",
    primaryClass = "hep-ph",
    reportNumber = "DESY-12-013, MZ-TH-12-07, LPSC-12019",
    doi = "10.1140/epjc/s10052-012-2082-2",
    journal = "Eur. Phys. J. C",
    volume = "72",
    pages = "2082",
    year = "2012"
}


%% file: ATLAS-useful.bib
@Article{Evans:2008zzb,
      author         = "Evans, Lyndon and Bryant, Philip",
      title          = "{LHC Machine}",
      journal        = "JINST",
      volume         = "3",
      pages          = "S08001",
      doi            = "10.1088/1748-0221/3/08/S08001",
      year           = "2008",
      SLACcitation   = "%%CITATION = JINST,3,S08001;%%",
}

@article{Dulat:2015mca,
      author         = "Dulat, Sayipjamal and Hou, Tie-Jiun and Gao, Jun and
                        Guzzi, Marco and Huston, Joey and Nadolsky, Pavel and
                        Pumplin, Jon and Schmidt, Carl and Stump, Daniel and Yuan,
                        C. P.",
      title          = "{New parton distribution functions from a global analysis
                        of quantum chromodynamics}",
      journal        = "Phys. Rev. D",
      volume         = "93",
      year           = "2016",
      number         = "3",
      pages          = "033006",
      doi            = "10.1103/PhysRevD.93.033006",
      eprint         = "1506.07443",
      archivePrefix  = "arXiv",
      primaryClass   = "hep-ph",
      SLACcitation   = "%%CITATION = ARXIV:1506.07443;%%"
}

@article{Ball:2012cx,
      author         = "{NNPDF Collaboration} and Ball, Richard D. and others",
      title          = "{Parton distributions with LHC data}",
      journal        = "Nucl. Phys. B",
      volume         = "867",
      year           = "2013",
      pages          = "244",
      doi            = "10.1016/j.nuclphysb.2012.10.003",
      eprint         = "1207.1303",
      archivePrefix  = "arXiv",
      primaryClass   = "hep-ph",
      reportNumber   = "EDINBURGH-2012-08, IFUM-FT-997, FR-PHENO-2012-014,
                        RWTH-TTK-12-25, CERN-PH-TH-2012-037, SFB-CPP-12-47\,
                        --CERN-PH-TH-2012-037",
      SLACcitation   = "%%CITATION = ARXIV:1207.1303;%%"
}

@article{Ball:2014uwa,
      author         = "{NNPDF Collaboration} and Ball, Richard D. and others",
      title          = "{Parton distributions for the LHC run II}",
      journal        = "JHEP",
      volume         = "04",
      year           = "2015",
      pages          = "040",
      doi            = "10.1007/JHEP04(2015)040",
      eprint         = "1410.8849",
      archivePrefix  = "arXiv",
      primaryClass   = "hep-ph",
      reportNumber   = "EDINBURGH-2014-15, IFUM-1034-FT, CERN-PH-TH-2013-253,
                        OUTP-14-11P, CAVENDISH-HEP-14-11",
      SLACcitation   = "%%CITATION = ARXIV:1410.8849;%%"
}

@Article{Sjostrand:2007gs,
	author    = "Sj{\"o}strand, T. and Mrenna, S. and Skands, P.",
	title     = "{A brief introduction to PYTHIA 8.1}",
	journal   = "Comput. Phys. Commun.",
	volume    = "178",
	year      = "2008",
	pages     = "852-867",
	eprint    = "0710.3820",
	archivePrefix = "arXiv",
	primaryClass  =  "hep-ph",
	doi       = "10.1016/j.cpc.2008.01.036",
	SLACcitation  = "%%CITATION = 0710.3820;%%"
}

@article{Sjostrand:2014zea,
      author         = "Sj{\"o}strand, Torbj{\"o}rn and Ask, Stefan and Christiansen,
                        Jesper R. and Corke, Richard and Desai, Nishita and Ilten,
                        Philip and Mrenna, Stephen and Prestel, Stefan and
                        Rasmussen, Christine O. and Skands, Peter Z.",
      title          = "{An introduction to PYTHIA 8.2}",
      journal        = "Comput. Phys. Commun.",
      volume         = "191",
      year           = "2015",
      pages          = "159",
      doi            = "10.1016/j.cpc.2015.01.024",
      eprint         = "1410.3012",
      archivePrefix  = "arXiv",
      primaryClass   = "hep-ph",
      reportNumber   = "LU-TP-14-36, MCNET-14-22, CERN-PH-TH-2014-190,
                        FERMILAB-PUB-14-316-CD, DESY-14-178, SLAC-PUB-16122,
                        --FERMILAB-PUB-14-316-CD",
      SLACcitation   = "%%CITATION = ARXIV:1410.3012;%%"
}

@article{Agostinelli:2002hh,
      author         = "Agostinelli, S. and others",
      title          = "{\textsc{Geant4} -- a simulation toolkit}",
      journal        = "Nucl. Instrum. Meth. A",
      volume         = "506",
      year           = "2003",
      pages          = "250",
      doi            = "10.1016/S0168-9002(03)01368-8",
      reportNumber   = "SLAC-PUB-9350, FERMILAB-PUB-03-339",
      SLACcitation   = "%%CITATION = NUIMA,A506,250;%%"
}

@Booklet{RooFit,
      author         = "Verkerke, Wouter and Kirkby, David",
      title          = "{The RooFit toolkit for data modeling}",
      year           = "2003",
      eprint         = "physics/0306116",
      archivePrefix  = "arXiv",
      primaryClass   = "physics.data-an",
}


%% file: ATLAS.bib
@Article{PERF-2007-01,
    author         = "{ATLAS Collaboration}",
    title          = "{The ATLAS Experiment at the CERN Large Hadron Collider}",
    journal        = "JINST",
    volume         = "3",
    year           = "2008",
    pages          = "S08003",
    doi            = "10.1088/1748-0221/3/08/S08003",
    primaryClass   = "hep-ex",
}

@Article{SOFT-2010-01,
    author         = "{ATLAS Collaboration}",
    title          = "{The ATLAS Simulation Infrastructure}",
    journal        = "Eur. Phys. J. C",
    volume         = "70",
    year           = "2010",
    pages          = "823",
    doi            = "10.1140/epjc/s10052-010-1429-9",
    eprint         = "1005.4568",
    archivePrefix  = "arXiv",
    primaryClass   = "physics.ins-det",
}

@Article{BPHY-2011-06,
    author         = "{ATLAS Collaboration}",
    title          = "{Measurement of upsilon production in \(7\,\text{TeV}\) \(pp\) collisions at ATLAS}",
    journal        = "Phys. Rev. D",
    volume         = "87",
    year           = "2013",
    pages          = "052004",
    doi            = "10.1103/PhysRevD.87.052004",
    reportNumber   = "CERN-PH-EP-2012-295",
    eprint         = "1211.7255",
    archivePrefix  = "arXiv",
    primaryClass   = "hep-ex",
}

@Article{BPHY-2013-04,
    author         = "{ATLAS Collaboration}",
    title          = "{Measurement of \(D^{*\pm}\), \(D^\pm\) and \(D_s^\pm\) meson production cross sections in \(pp\) collisions at \(\sqrt{s} = 7\,\text{TeV}\) with the ATLAS detector}",
    journal        = "Nucl. Phys. B",
    volume         = "907",
    year           = "2016",
    pages          = "717",
    doi            = "10.1016/j.nuclphysb.2016.04.032",
    reportNumber   = "CERN-PH-EP-2015-288",
    eprint         = "1512.02913",
    archivePrefix  = "arXiv",
    primaryClass   = "hep-ex",
}

@Article{PERF-2015-08,
    author         = "{ATLAS Collaboration}",
    title          = "{Performance of the ATLAS track reconstruction algorithms in dense environments in LHC Run~2}",
    journal        = "Eur. Phys. J. C",
    volume         = "77",
    year           = "2017",
    pages          = "673",
    doi            = "10.1140/epjc/s10052-017-5225-7",
    reportNumber   = "CERN-EP-2017-045",
    eprint         = "1704.07983",
    archivePrefix  = "arXiv",
    primaryClass   = "hep-ex",
}

@Article{TRIG-2016-01,
    author         = "{ATLAS Collaboration}",
    title          = "{Performance of the ATLAS trigger system in 2015}",
    journal        = "Eur. Phys. J. C",
    volume         = "77",
    year           = "2017",
    pages          = "317",
    doi            = "10.1140/epjc/s10052-017-4852-3",
    reportNumber   = "CERN-EP-2016-241",
    eprint         = "1611.09661",
    archivePrefix  = "arXiv",
    primaryClass   = "hep-ex",
}

@Article{DAPR-2018-01,
    author         = "{ATLAS Collaboration}",
    title          = "{ATLAS data quality operations and performance for 2015--2018 data-taking}",
    journal        = "JINST",
    volume         = "15",
    year           = "2020",
    pages          = "P04003",
    doi            = "10.1088/1748-0221/15/04/P04003",
    reportNumber   = "CERN-EP-2019-207",
    eprint         = "1911.04632",
    archivePrefix  = "arXiv",
    primaryClass   = "physics.ins-det",
}

@Article{MUON-2018-03,
    author         = "{ATLAS Collaboration}",
    title          = "{Muon reconstruction and identification efficiency in ATLAS using the full Run~2 \(pp\) collision data set at \(\sqrt{s} = 13\,\text{TeV}\)}",
    journal        = "Eur. Phys. J. C",
    volume         = "81",
    year           = "2021",
    pages          = "578",
    doi            = "10.1140/epjc/s10052-021-09233-2",
    reportNumber   = "CERN-EP-2020-199",
    eprint         = "2012.00578",
    archivePrefix  = "arXiv",
    primaryClass   = "hep-ex",
}

@Article{PIX-2018-001,
    author         = "Abbott, B. and others",
    title          = "{Production and integration of the ATLAS Insertable B-Layer}",
    journal        = "JINST",
    volume         = "13",
    year           = "2018",
    pages          = "T05008",
    doi            = "10.1088/1748-0221/13/05/T05008",
    eprint         = "1803.00844",
    archivePrefix  = "arXiv",
    primaryClass   = "physics.ins-det",
}

@Article{TRIG-2018-01,
    author         = "{ATLAS Collaboration}",
    title          = "{Performance of the ATLAS muon triggers in Run~2}",
    journal        = "JINST",
    volume         = "15",
    year           = "2020",
    pages          = "P09015",
    doi            = "10.1088/1748-0221/15/09/p09015",
    reportNumber   = "CERN-EP-2020-031",
    eprint         = "2004.13447",
    archivePrefix  = "arXiv",
    primaryClass   = "physics.ins-det",
}

@Article{DAPR-2021-01,
    author         = "{ATLAS Collaboration}",
    title          = "{Luminosity determination in \(pp\) collisions at \(\sqrt{s} = 13\,\text{TeV}\) using the ATLAS detector at the LHC}",
    journal        = "Eur. Phys. J. C",
    volume         = "83",
    year           = "2023",
    pages          = "982",
    doi            = "10.1140/epjc/s10052-023-11747-w",
    reportNumber   = "CERN-EP-2022-281",
    eprint         = "2212.09379",
    archivePrefix  = "arXiv",
    primaryClass   = "hep-ex",
}

@Article{MUON-2022-01,
    author         = "{ATLAS Collaboration}",
    title          = "{Studies of the muon momentum calibration and performance of the ATLAS detector with \(pp\) collisions at \(\sqrt{s} = 13\,\text{TeV}\)}",
    journal        = "Eur. Phys. J. C",
    volume         = "83",
    year           = "2023",
    pages          = "686",
    doi            = "10.1140/epjc/s10052-023-11584-x",
    reportNumber   = "CERN-EP-2022-274",
    eprint         = "2212.07338",
    archivePrefix  = "arXiv",
    primaryClass   = "hep-ex",
}

@Article{SOFT-2022-02,
    author         = "{ATLAS Collaboration}",
    title          = "{Software and computing for Run~3 of the ATLAS experiment at the LHC}",
    year           = "2024",
    reportNumber   = "CERN-EP-2024-100",
    eprint         = "2404.06335",
    archivePrefix  = "arXiv",
    primaryClass   = "hep-ex",
}

@Booklet{ATL-SOFT-PUB-2025-001,
    author         = "{ATLAS Collaboration}",
    title          = "{ATLAS Computing Acknowledgements}",
    howpublished   = "{ATL-SOFT-PUB-2025-001}",
    url            = "https://cds.cern.ch/record/2922210",
    year           = "2025",
}

@Report{ATLAS-TDR-19,
    author         = "{ATLAS Collaboration}",
    title          = "{ATLAS Insertable B-Layer: Technical Design Report}",
    type           = "ATLAS-TDR-19; CERN-LHCC-2010-013",
    year           = "2010",
    url            = "https://cds.cern.ch/record/1291633",
    related        = "ATLAS-TDR-19-addm",
    relatedstring  = "Addendum:",
}

@Article{LUCID2,
    author         = "Avoni, G. and others",
    title          = "{The new LUCID-2 detector for luminosity measurement and monitoring in ATLAS}",
    journal        = "JINST",
    volume         = "13",
    year           = "2018",
    pages          = "P07017",
    doi            = "10.1088/1748-0221/13/07/P07017",
}


%% file: CMS.bib
@Article{CMS-BPH-18-003,
    author         = "{CMS Collaboration}",
    title          = "{Measurement of prompt open-charm production cross sections in proton--proton collisions at \(\sqrt{s} = 13\,\text{TeV}\)}",
    journal        = "JHEP",
    volume         = "11",
    year           = "2021",
    pages          = "225",
    doi            = "10.1007/JHEP11(2021)225",
    reportNumber   = "CERN-EP-2021-085",
    eprint         = "2107.01476",
    archivePrefix  = "arXiv",
    primaryClass   = "hep-ex",
}


%% file: PubNotes.bib
@Booklet{ATL-PHYS-PUB-2014-021,
    author         = "{ATLAS Collaboration}",
    title          = "{ATLAS Pythia~8 tunes to \(7~\text{TeV}\) data}",
    howpublished   = "{ATL-PHYS-PUB-2014-021}",
    url            = "https://cds.cern.ch/record/1966419",
    year           = "2014",
}

@Booklet{ATL-PHYS-PUB-2015-026,
    author         = "{ATLAS Collaboration}",
    title          = "{Vertex Reconstruction Performance of the ATLAS Detector at \(\sqrt{s} = 13~\text{TeV}\)}",
    howpublished   = "{ATL-PHYS-PUB-2015-026}",
    url            = "https://cds.cern.ch/record/2037717",
    year           = "2015",
}

@Booklet{ATL-PHYS-PUB-2015-051,
    author         = "{ATLAS Collaboration}",
    title          = "{Early Inner Detector Tracking Performance in the 2015 Data at \(\sqrt{s} = 13~\text{TeV}\)}",
    howpublished   = "{ATL-PHYS-PUB-2015-051}",
    url            = "https://cds.cern.ch/record/2110140",
    year           = "2015",
}

@Booklet{ATL-PHYS-PUB-2016-017,
    author         = "{ATLAS Collaboration}",
    title          = "{The Pythia~8 A3 tune description of ATLAS minimum bias and inelastic measurements incorporating the Donnachie--Landshoff diffractive model}",
    howpublished   = "{ATL-PHYS-PUB-2016-017}",
    url            = "https://cds.cern.ch/record/2206965",
    year           = "2016",
}

@Booklet{ATL-PHYS-PUB-2018-032,
    author         = "{ATLAS Collaboration}",
    title          = "{Prospects for lepton flavour violation measurements in \(\tau \to 3 \mu\) decays with the ATLAS detector at the HL-LHC}",
    howpublished   = "{ATL-PHYS-PUB-2018-032}",
    url            = "https://cds.cern.ch/record/2647956",
    year           = "2018",
}
